\documentclass[10pt,twocolumn,letterpaper,pagebackref,breaklinks,colorlinks,allcolors=cvprblue]{article}
\usepackage[pagenumbers]{cvpr}      
\usepackage[table]{xcolor}

\usepackage{orcidlink}

\definecolor{cvprblue}{rgb}{0.21,0.49,0.74}

\def\confName{CVPR}
\def\confYear{2027}

\title{FlowSlide-GS: Residual-Guided Periodic Gaussian Transport\\for 4D CT Reconstruction}

\author{
Kangneng Zhou$^{1}$ \quad
Tao Wu$^{2}$ \quad
Xiangyue Liu$^{3}$ \quad
Jie Zhang$^{4}$\orcidlink{0000-0001-8219-5590} \quad
Chubin Ou$^{5}$ \\
Gadeng Luosang$^{6}$ \quad
Weili Jiang$^{7*}$ \\
$^{1}$Nankai University 
\quad $^{2}$Computer Vision Center \\
\quad $^{3}$The Affiliated Cancer Hospital of Zhengzhou University \& Henan Cancer Hospital \\
\quad $^{4}$ Macao Polytechnic University 
\quad $^{5}$ Shenzhen research institute of big data 
\quad $^{6}$ Tibet University
\\
\quad $^{7}$  Southwest Jiaotong University \\
\\
{\tt\small jiangweili@swjtu.edu.cn}
}

\begin{document}
\maketitle
\begin{abstract}
Sparse-view four-dimensional computed tomography (4D CT) pools observations across respiratory phases to constrain motion under limited angular coverage. This coupling can propagate a local correction to other phases and degrade their reconstructions. We introduce FlowSlide-GS, which uses shared periodic Gaussian transport to estimate respiratory motion and controls where local corrections take effect. Projection-loss gradients select Gaussian primitives for directional correction. A masked periodic residual dictionary then refines selected phase centers with the shared model frozen, preserving the reconstructed volumes at inactive centers. A separate calibration step uses same-phase training residuals to refine normalized projections with the volumes held fixed. Across 13 public and 10 private cases evaluated with prepared or simulated projections, FlowSlide-GS improves mean volumetric PSNR by 1.65 and 0.59~dB, respectively, over our reproduction of $X^2$-Gaussian.
\end{abstract}    
\section{Introduction}

Sparse-view 4D CT aims to recover a 3D attenuation volume for each respiratory phase from a limited number of X-ray projections. Because the acquisition angle varies with respiratory state, each phase has limited angular coverage, making accurate reconstruction highly dependent on information shared across phases. Existing methods therefore commonly employ shared motion representations to jointly constrain reconstruction across respiratory phases~\cite{reed2021dynamicct,shao2023stinr,shao2024dreme,yu2025x2gaussian}. However, global information sharing does not necessarily imply that local deformation corrections should also be shared globally. Consider the relative sliding motion between the lung and chest wall along their interface during respiration. When the shared motion representation fails to capture a local displacement at a particular phase, updating shared parameters using the corresponding projection residuals may improve the target region but simultaneously alter other regions or respiratory phases. This introduces a fundamental challenge: \emph{how can cross-phase observations be exploited to constrain global respiratory motion while restricting additional deformation corrections to only the spatial regions and phase centers that require refinement?}

\begin{figure}[t]
    \centering
    \includegraphics[width=0.84\linewidth]{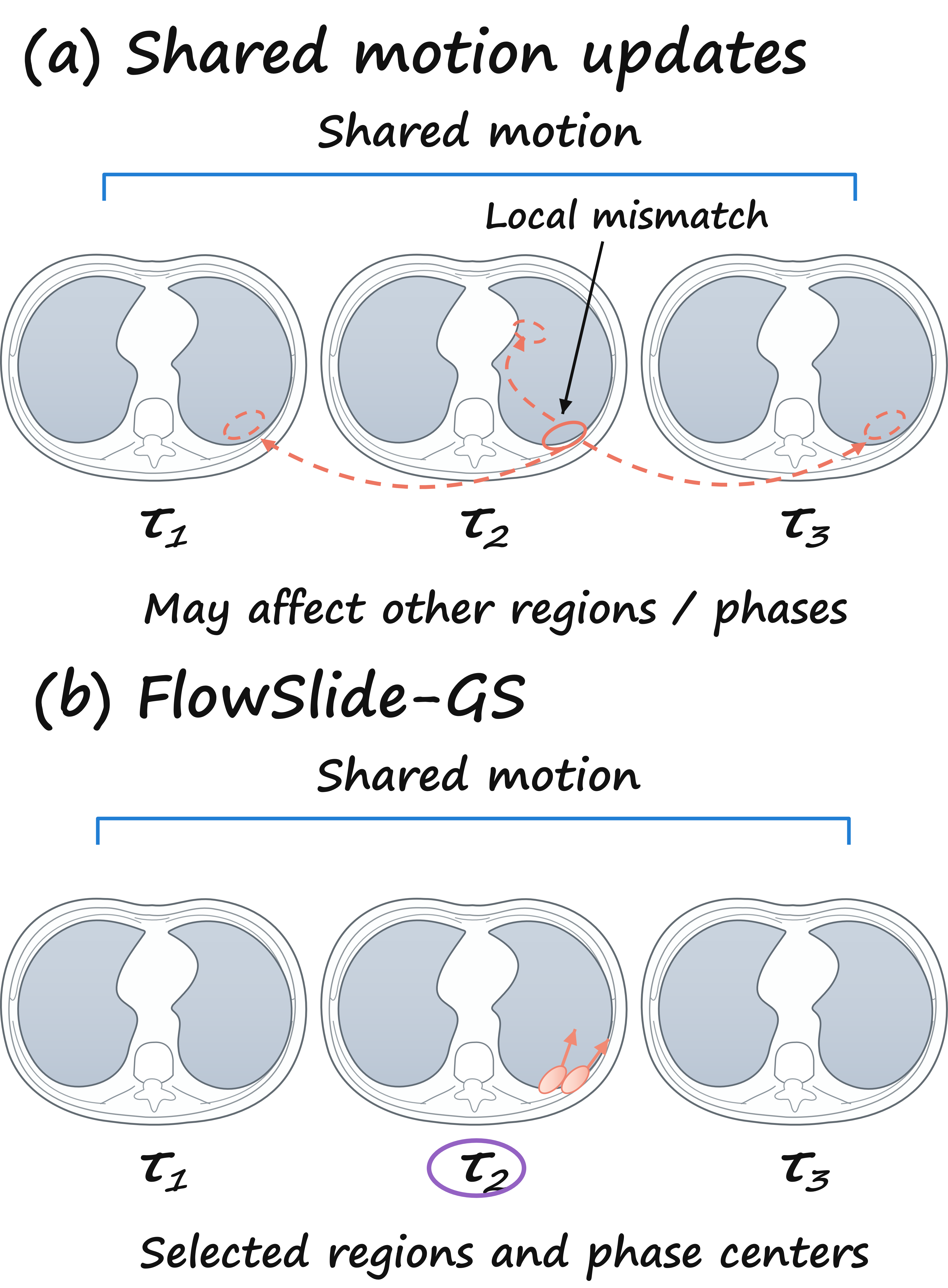}
    \caption{
    Motivation of FlowSlide-GS.
    (a) Shared motion updates may propagate local corrections to
    unintended regions and respiratory phases.
    (b) FlowSlide-GS restricts additional deformation corrections
    to selected spatial regions and phase centers while preserving
    shared motion across phases.
    }
    \label{fig:introduction}
\end{figure}

Prior work has improved sparse-view 4D CT reconstruction from several complementary perspectives. Joint reconstruction and motion-estimation methods integrate sparse observations from different respiratory phases through shared reference volumes and motion representations~\cite{reed2021dynamicct,shao2023stinr,shao2024dreme}. Radiative Gaussian representations further enable differentiable modeling for transmission imaging~\cite{cai2024xgaussian,zha2024r2gaussian}, while $X^2$-Gaussian extends this paradigm to continuous-phase reconstruction through a shared dynamic representation~\cite{yu2025x2gaussian}. Meanwhile, studies of sliding motion demonstrate that tangential and normal displacements at tissue interfaces can be modeled under different constraints, providing a basis for more anatomically appropriate local deformation~\cite{schmidtrichberg2012slipping}. Projection-domain completion offers another complementary strategy by directly estimating missing projections in measurement space~[2]. Despite these advances, an important question remains insufficiently addressed: \emph{when shared motion representations exhibit local mismatches, how can additional deformation capacity be allocated to the relevant spatial regions without introducing unintended updates to other regions or respiratory phases?} In particular, spatial localization alone is insufficient if the resulting corrections remain coupled across respiratory phases.

To address this challenge, we propose \textbf{FlowSlide-GS}, a residual-guided local Gaussian transport framework for sparse-view 4D CT reconstruction. Our key insight is to preserve cross-phase sharing for globally consistent motion modeling while introducing additional deformation only where and when the shared representation is insufficient. Specifically, projection-adjoint gradients guide the allocation of local deformation capacity, while Jacobian-coupled Gaussian transport and a masked periodic residual dictionary enable spatially and phase-specific corrections without affecting inactive phase centers. Furthermore, we decouple 3D volume reconstruction from 2D projection calibration by refining normalized projections using training-view residuals while keeping the reconstructed volume fixed, enabling independent evaluation of volumetric reconstruction and held-out-view prediction.

Our main contributions are:
\begin{itemize}
    \item A residual-guided Gaussian transport framework that combines periodic Jacobian-coupled motion with adjoint-guided directional corrections.

    \item A masked phase-residual dictionary that refines difficult phases while keeping inactive phase centers unchanged.

    \item An explicit separation of 3D reconstruction and fixed-volume 2D projection calibration, evaluated on public and private data.
\end{itemize}

\section{Related Work}
\label{sec:related}

\subsection{Sparse-View CT Reconstruction and Dynamic Attenuation Fields}

Classical CT reconstruction uses analytical inversion or iterative projection consistency, including FDK~\cite{feldkamp1984fdk}, SART~\cite{andersen1984sart}, and total-variation regularization~\cite{sidky2008tv}; TIGRE provides practical CBCT operators~\cite{biguri2016tigre}. Sparse-view 4D CT remains challenging because each respiratory phase has limited angular coverage. CTNet addresses incomplete measurements through sinogram completion~\cite{anirudh2018losetheviews}, whereas neural attenuation fields optimize continuous representations directly from projections. Representative methods include NAF~\cite{zha2022naf}, NeAT~\cite{rueckert2022neat}, and SAX-NeRF~\cite{cai2024saxnerf}. Dynamic CT methods further jointly estimate anatomy and motion~\cite{reed2021dynamicct}, with STINR and DREME extending implicit reconstruction to dynamic CBCT~\cite{shao2023stinr,shao2024dreme}.

These methods establish projection-driven reconstruction and cross-phase information sharing. However, shared motion parameters couple updates across regions and phases, so local residuals may alter otherwise well-reconstructed content. Discrepancy-aware attenuation fields allocate capacity according to frequency-dependent errors~\cite{oh2026ghnaf}, but do not explicitly determine the spatial, phase, and representation-domain scope of each correction. FlowSlide-GS addresses this issue by assigning residual corrections to their corresponding scopes.

\begin{figure*}[ht]
    \centering
    \includegraphics[width=1.0 \linewidth]
    {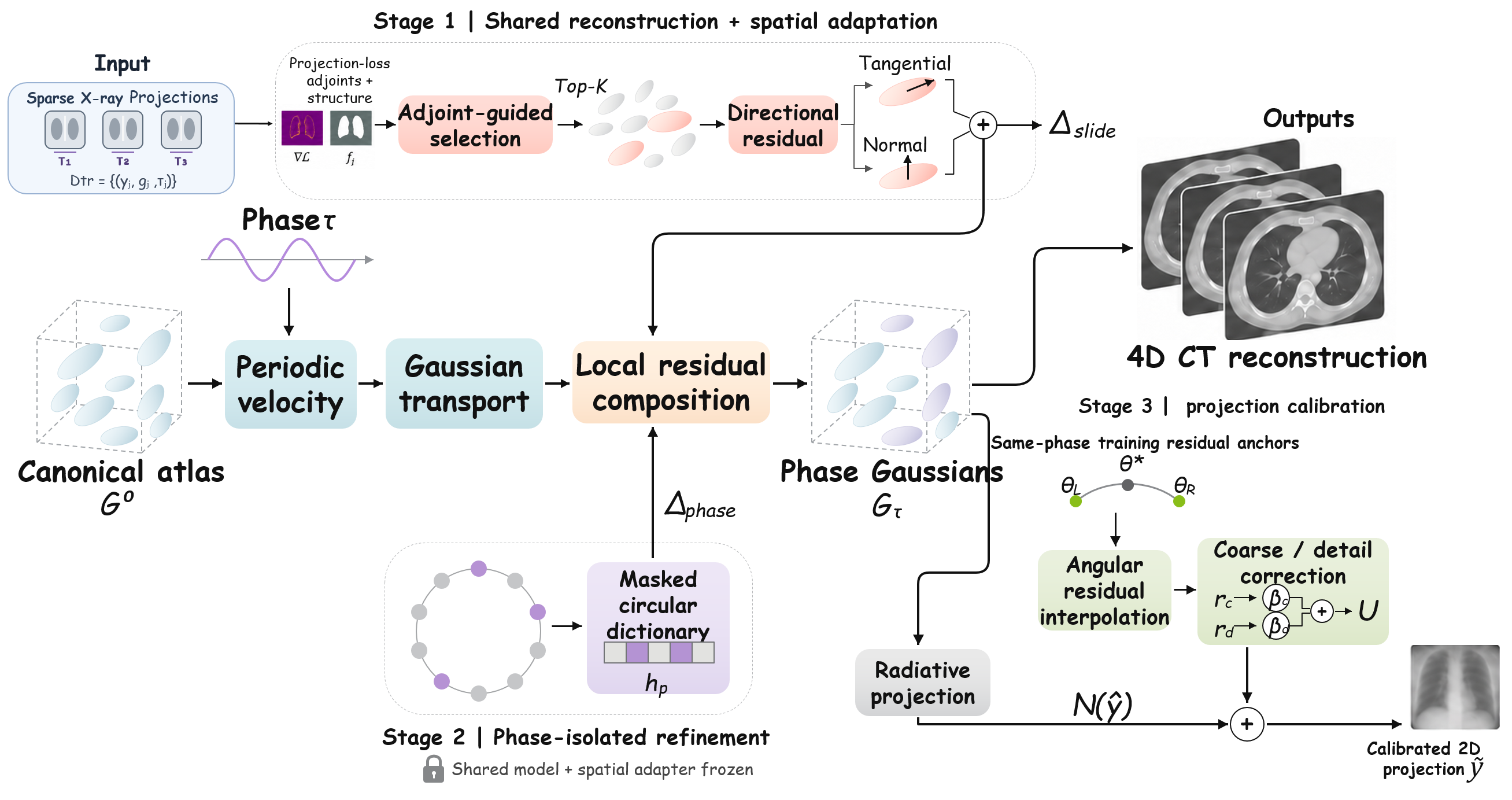}
    \caption{
    Overview of FlowSlide-GS.
    Stage~1 learns shared periodic Gaussian transport and applies
    adjoint-guided local spatial corrections.
    Stage~2 freezes the shared model and refines selected phase centers.
    Stage~3 fixes the reconstructed 3D Gaussians and calibrates normalized
    projections using same-phase residuals.
    Ours-Recon outputs phase-wise attenuation volumes, while Ours-Full
    changes only the 2D projection predictions.
    }
    \label{fig:overview}
\end{figure*}

\subsection{Dynamic Gaussian Representations and Local Motion}

Dynamic neural fields represent temporal variation through deformation or scene flow. D-NeRF, Nerfies, and non-rigid NeRF use deformation fields~\cite{pumarola2021dnerf,park2021nerfies,tretschk2021nrnerf}; neural scene flow and HyperNeRF model spatiotemporal and structural changes~\cite{li2021nsff,park2021hypernerf}; and HexPlane and K-Planes provide compact plane factorizations~\cite{cao2023hexplane,fridovichkeil2023kplanes}. These representations enable temporal sharing but do not constrain local updates to particular regions or time points.

Gaussian representations provide explicit and differentiable primitives. Dynamic 3D Gaussians, 4D Gaussian Splatting, and deformable Gaussians extend 3D Gaussian Splatting to time-varying scenes~\cite{kerbl2023gaussian,luiten2024dynamic,wu2024fourdimensional,yang2024deformable}; SC-GS introduces sparse motion controls~\cite{huang2024scgs}, and Mip-Splatting addresses aliasing~\cite{yu2024mipsplatting}. For transmission imaging, X-Gaussian supports X-ray novel-view synthesis~\cite{cai2024xgaussian}, R$^2$-Gaussian performs tomographic reconstruction~\cite{zha2024r2gaussian}, and $X^2$-Gaussian models continuous respiratory motion~\cite{yu2025x2gaussian}. Direction-dependent registration further distinguishes tangential and normal motion near sliding interfaces~\cite{schmidtrichberg2012slipping}. Building on these foundations, FlowSlide-GS focuses on allocating residual corrections within a shared dynamic Gaussian representation rather than introducing dynamic Gaussians themselves.

\subsection{Phase-Isolated and Measurement-Space Correction}

Cross-phase sharing improves sparse-view constraints but can propagate a phase-specific update to other phases. FlowSlide-GS freezes the shared model and uses a masked periodic residual dictionary whose updates are zero at inactive phase centers, thereby restricting phase refinement.

Measurement-space correction is also established in CT, for example through sinogram completion~\cite{anirudh2018losetheviews}. FlowSlide-GS instead calibrates held-out projections after reconstruction by interpolating same-phase training residuals and fitting two ridge-regularized gains~\cite{hoerl1970ridge}. This correction leaves the Gaussian geometry and attenuation volume fixed. Accordingly, we evaluate 3D reconstruction and 2D projection prediction separately.

\section{Method}
\label{sec:method}

\subsection{Problem Setting and Overview}
\label{sec:problem}

We reconstruct a periodic 4D attenuation field from
\begin{equation}
\mathcal{D}_{\mathrm{tr}}
=
\{(y_j,g_j,\tau_j)\}_{j=1}^{M},
\end{equation}
where $y_j$ is a training projection, $g_j$ is its acquisition geometry,
and $\tau_j\in[0,1)$ is its respiratory phase. Ground-truth volumes and
held-out projections are used only for evaluation.

Sparse-view 4D CT requires observations to be shared across phases, but a
local residual should not alter unrelated regions or phases. Moreover,
remaining projection errors do not necessarily require changes to the
reconstructed volume. FlowSlide-GS therefore assigns residual corrections
to three scopes, as illustrated in Fig.~\ref{fig:overview}: Stage~1 selects
where spatial correction is needed, Stage~2 restricts which phase centers
can be refined, and Stage~3 corrects only the normalized projection after
the 3D representation is fixed.

\subsection{Stage 1: Shared Reconstruction and Spatial Adaptation}
\label{sec:spatial}

\paragraph{Shared periodic Gaussian transport.}
We represent the attenuation volume using radiative Gaussians, each
parameterized by a center, covariance, and attenuation amplitude.
Differentiable transmission rendering follows
R$^2$-Gaussian and $X^2$-Gaussian
~\cite{zha2024r2gaussian,yu2025x2gaussian}.

Let $G_0$ denote the learned canonical Gaussian atlas. The shared periodic
map $\Phi_\tau$ models global respiratory motion, while the local map
$\Psi_\tau$ introduces residual-guided spatial correction:
\begin{equation}
G_\tau
=
\operatorname{Transport}
\left(G_0,\Psi_\tau\circ\Phi_\tau\right).
\end{equation}
The periodic basis, velocity integration, and Jacobian propagation follow
continuous dynamic representations and $X^2$-Gaussian
~\cite{sitzmann2020siren}.

\paragraph{Adjoint-guided spatial support.}
Backpropagation through the renderer gives gradients with respect to each
transported center $m_{i,\tau}$ and attenuation amplitude $a_{i,\tau}$.
We map the center gradient to the canonical coordinates:
\begin{equation}
q_i
=
J_{i,\tau}^{\top}
\frac{\partial L_{\mathrm{proj}}}{\partial m_{i,\tau}},
\qquad
p_i
=
\frac{\partial L_{\mathrm{proj}}}{\partial a_{i,\tau}}.
\end{equation}
For visible primitives, the normalized residual evidence is
\begin{equation}
e_i
=
\frac{1}{2}
\left[
\frac{\|q_i\|^2}
{\operatorname{mean}_{\mathrm{vis}}\|q\|^2+\epsilon}
+
\frac{|p_i|}
{\operatorname{mean}_{\mathrm{vis}}|p|+\epsilon}
\right].
\end{equation}
We maintain global and phase-conditioned exponential moving averages of this
evidence. It measures sensitivity to the current projection discrepancy and
requires neither anatomical labels nor an additional renderer backward pass.

The selector combines residual evidence with structural features, including
log attenuation, anisotropy, and relative Gaussian size. It assigns bounded
scores $s_\eta(f_i,\tau)$ and retains a fixed-capacity support:
\begin{equation}
S_\tau
=
\operatorname{TopK}_i
\left[
s_\eta(f_i,\tau)
+
\lambda_e\bar e_{i,\tau}
\right].
\end{equation}
The selector is trained by ranking high-evidence primitives above
low-evidence primitives:
\begin{equation}
L_{\mathrm{rank}}
=
\operatorname{mean}_{(h,l)}
\operatorname{softplus}
\left(
-\frac{s_h-s_l}{T}
\right).
\end{equation}
Thus, primitives outside the hard support still supervise the selector.

\paragraph{Directional local correction.}
We use the shortest covariance axis as a canonical normal proxy and transport
it with the inverse-transpose Jacobian:
\begin{equation}
n_{i,\tau}
=
\operatorname{normalize}
\left(
J_{i,\tau}^{-\top}n_i^0
\right).
\end{equation}
For a selected primitive, the spatial residual is decomposed into tangential
and normal components:
\begin{equation}
\begin{aligned}
\Delta x_{i,\tau}^{\mathrm{slide}}
={}&
\gamma_{i,\tau}^{t}
\left(
I-n_{i,\tau}n_{i,\tau}^{\top}
\right)u_\psi \\
&+
\gamma_{i,\tau}^{n}
b_\psi n_{i,\tau}.
\end{aligned}
\end{equation}
The tangential branch represents sliding-like motion, whereas the separately
gated normal branch handles departures that cannot be expressed tangentially.
The gates depend jointly on structural and residual cues.

Let $\Delta_\tau$ denote the combined local residual. Its Jacobian is composed
with the global transport:
\begin{equation}
J_{i,\tau}^{\mathrm{total}}
=
\left(
I+\nabla_x\Delta_\tau
\right)
J_{i,\tau}^{\mathrm{global}}.
\end{equation}
The composed Jacobian transports Gaussian centers, covariances, and
attenuation amplitudes.

\subsection{Stage 2: Phase-Isolated Refinement}
\label{sec:phase}

After Stage~1, we freeze the shared reconstruction model and spatial adapter.
A masked circular dictionary then introduces residual capacity only for
selected phase centers:
\begin{equation}
\Delta_\tau^{\mathrm{phase}}(x)
=
g(x,\tau)b_{\max}
\tanh
\left[
\sum_{p=0}^{P-1}
h_pK_p(\tau)c_p(x)
\right],
\end{equation}
where $h_p\in\{0,1\}$ activates phase coefficient $p$, $c_p(x)$ is predicted
by a shared spatial network, and $K_p$ interpolates the coefficients on the
circular phase domain.

For adjacent phase centers, we use raised-cosine interpolation:
\begin{equation}
\alpha(u)
=
\frac{1}{2}
\left[
1-\cos(\pi u)
\right],
\qquad u\in[0,1],
\end{equation}
with weights $1-\alpha$ and $\alpha$. The resulting cardinal property,
$K_p(\tau_q)=\delta_{pq}$, gives
\begin{equation}
h_q=0
\quad\Longrightarrow\quad
\Delta_{\tau_q}^{\mathrm{phase}}(x)=0.
\end{equation}
Thus, inactive phase centers remain unchanged even though all coefficients
share the same spatial network. Corrections can still interpolate between
neighboring active centers.

We identify difficult phases using the moving average $L_p$ of the normalized
projection-shape loss:
\begin{equation}
z_p
=
\frac{
\log L_p-\operatorname{median}(\log L)
}{
1.4826\operatorname{MAD}(\log L)+\epsilon
}.
\end{equation}
Phases whose scores exceed the prescribed threshold form the active set.

\subsection{Stage 3: Geometry-Fixed Projection Calibration}
\label{sec:calibration}

After reconstructing the phase Gaussians, we freeze the complete 3D model and
correct only the remaining normalized projection residual. Define
\begin{equation}
N(y)
=
\frac{y}
{\max(\max_u y(u),\epsilon)}.
\end{equation}
For training view $j$, the residual anchor is
\begin{equation}
r_j
=
N(y_j)-N(\widehat y_j).
\end{equation}
Held-out targets are not used to construct calibrated predictions.

\begin{table}[ht]
\centering
\small
\caption{Public benchmark results (PSNR in dB). Overall means weight the 13 cases equally.}
\label{tab:public}
\resizebox{0.99\linewidth}{!}{%
\begin{tabular}{@{}llrrrr@{}}
\toprule
Subset & Method & 2D PSNR & 2D SSIM & 3D PSNR & 3D SSIM\\
\midrule
DIR & FDK & 23.214 & 0.836 & 26.652 & 0.512\\
DIR & NAF & 41.438 & 0.942 & 36.327 & 0.898\\
DIR & SAX-NeRF & 42.113 & 0.984 & 37.592 & 0.916\\
DIR & R$^2$-Gaussian & 43.822 & 0.987 & 37.069 & 0.915\\
DIR & $X^2$-Gaussian & 43.997 & 0.990 & 39.288 & 0.954\\
DIR & Ours-Recon & 43.653 & 0.990 & \cellcolor[gray]{0.90}\textbf{40.763} & \cellcolor[gray]{0.90}\textbf{0.965}\\
DIR & Ours-Full & \cellcolor[gray]{0.90}\textbf{46.359} & \cellcolor[gray]{0.90}\textbf{0.992} & \cellcolor[gray]{0.90}\textbf{40.763} & \cellcolor[gray]{0.90}\textbf{0.965}\\
\midrule
4DLung & FDK & 24.612 & 0.746 & 21.362 & 0.203\\
4DLung & NAF & 37.143 & 0.785 & 33.509 & 0.889\\
4DLung & SAX-NeRF & 45.480 & 0.992 & 33.878 & 0.895\\
4DLung & R$^2$-Gaussian & 47.176 & 0.994 & 34.513 & 0.912\\
4DLung & $X^2$-Gaussian & 48.690 & 0.997 & 38.606 & 0.957\\
4DLung & Ours-Recon & 48.845 & 0.996 & \cellcolor[gray]{0.90}\textbf{39.844} & \cellcolor[gray]{0.90}\textbf{0.962}\\
4DLung & Ours-Full & \cellcolor[gray]{0.90}\textbf{50.295} & \cellcolor[gray]{0.90}\textbf{0.997} & \cellcolor[gray]{0.90}\textbf{39.844} & \cellcolor[gray]{0.90}\textbf{0.962}\\
\midrule
SPARE & FDK & 24.917 & 0.928 & 16.779 & 0.390\\
SPARE & NAF & 34.456 & 0.851 & 27.263 & 0.791\\
SPARE & SAX-NeRF & 44.019 & 0.991 & 27.823 & 0.815\\
SPARE & R$^2$-Gaussian & 46.340 & 0.994 & 27.683 & 0.840\\
SPARE & $X^2$-Gaussian & 48.657 & 0.998 & 32.284 & 0.922\\
SPARE & Ours-Recon & 47.772 & 0.998 & \cellcolor[gray]{0.90}\textbf{34.890} & \cellcolor[gray]{0.90}\textbf{0.937}\\
SPARE & Ours-Full & \cellcolor[gray]{0.90}\textbf{51.789} & \cellcolor[gray]{0.90}\textbf{0.998} & \cellcolor[gray]{0.90}\textbf{34.890} & \cellcolor[gray]{0.90}\textbf{0.937}\\
\midrule
Overall & FDK & 24.145 & 0.823 & 22.339 & 0.365\\
Overall & NAF & 38.175 & 0.861 & 33.151 & 0.870\\
Overall & SAX-NeRF & 43.848 & 0.989 & 33.909 & 0.885\\
Overall & R$^2$-Gaussian & 45.693 & 0.991 & 33.920 & 0.896\\
Overall & $X^2$-Gaussian & 46.877 & 0.994 & 37.409 & 0.948\\
Overall & Ours-Recon & 46.600 & 0.994 & \cellcolor[gray]{0.90}\textbf{39.054} & \cellcolor[gray]{0.90}\textbf{0.957}\\
Overall & Ours-Full & \cellcolor[gray]{0.90}\textbf{49.126} & \cellcolor[gray]{0.90}\textbf{0.995} & \cellcolor[gray]{0.90}\textbf{39.054} & \cellcolor[gray]{0.90}\textbf{0.957}\\
\bottomrule
\end{tabular}%
}
\end{table}

For a query angle $\vartheta$, we select its left and right angular neighbors
within the same phase. With circular distances $d_L$ and $d_R$, the interpolated
residual is
\begin{equation}
\bar r_{\tau,\vartheta}
=
\frac{
d_Rr_L+d_Lr_R
}{
d_L+d_R+\epsilon
}.
\end{equation}
No correction is applied when same-phase angular support is unavailable.

A local averaging operator $S$ separates coarse and detail residuals:
\begin{equation}
r_c=S(\bar r),
\qquad
r_d=\bar r-r_c.
\end{equation}
For each training anchor, leave-one-out interpolation excludes that anchor
when constructing its residual features. After resizing both components to
the detector grid, the feature matrix $F_j^{\mathrm{LOO}}$ is used to estimate
two phase-specific ridge gains~\cite{hoerl1970ridge}:
\begin{equation}
\beta_\tau
=
\left[
\sum_j F_j^{\top}F_j
+
\lambda s_\tau I
\right]^{-1}
\sum_j F_j^{\top}r_j.
\end{equation}
The two entries of $\beta_\tau$ are the coarse and detail gains
$\beta_{\tau,c}$ and $\beta_{\tau,d}$.

The calibrated projection is
\begin{equation}
\widetilde y
=
\operatorname{clip}_{[0,1]}
\left\{
N(\widehat y)
+
U\left(
\beta_{\tau,c}r_c+
\beta_{\tau,d}r_d
\right)
\right\},
\end{equation}
where $U$ denotes bilinear upsampling. The calibrator uses only training
residuals from the corresponding phase and does not modify the reconstructed
Gaussians. Consequently, Ours-Recon and Ours-Full produce identical volumes.

\subsection{Optimization and Staged Training}
\label{sec:optimization}

We use Adam~\cite{kingma2015adam} to optimize the canonical atlas, shared
motion, and residual adapters from training projections. Reference CT volumes
are used only for evaluation. The projection objective is
\begin{equation}
\begin{aligned}
L_{\mathrm{data}} ={}&
L_1
+\lambda_{\mathrm{mse}}L_{\mathrm{mse}}
+\lambda_{\mathrm{wav}}L_{\mathrm{wav}}\\
&+\lambda_{\mathrm{ssim}}(1-\mathrm{SSIM})
+\lambda_{\mathrm{rel}}L_{\mathrm{rel}}
+\lambda_{\mathrm{shape}}L_{\mathrm{shape}}.
\end{aligned}
\end{equation}
The structural term follows SSIM~\cite{wang2004ssim}, and the wavelet term
uses the one-level db1 formulation employed by $X^2$-Gaussian. The normalized
losses are
\begin{equation}
L_{\mathrm{rel}}
=
\operatorname{MSE}
\left(
\widehat y/s_y,y/s_y
\right),
\qquad
L_{\mathrm{shape}}
=
\operatorname{MSE}
\left(
\widehat y/s_{\widehat y},y/s_y
\right),
\end{equation}
where
\begin{equation}
s_y=\max(s_g(\max y),\epsilon),
\qquad
s_{\widehat y}
=
\max(s_g(\max\widehat y),0.1s_y,\epsilon).
\end{equation}

\section{Experiments}
\label{sec:experiments}

\begin{table}[t]
\centering
\caption{Results on ten private clinical 4D CT cases with simulated
projections.}
\label{tab:private}
\footnotesize
\setlength{\tabcolsep}{3pt}
\begin{tabular*}{\columnwidth}{@{\extracolsep{\fill}}lrrrr@{}}
\toprule
 & \multicolumn{2}{c}{2D Projection}
 & \multicolumn{2}{c}{3D Volume} \\
\cmidrule(lr){2-3}\cmidrule(lr){4-5}
Method
 & PSNR (dB) & SSIM
 & PSNR (dB) & SSIM \\
\midrule
FDK                       & 23.469 & 0.710822 & 31.588 & 0.474438 \\
NAF                       & 33.010 & 0.759849 & 44.153 & 0.969660 \\
SAX-NeRF                  & 46.353 & 0.989075 & 44.894 & 0.973383 \\
R$^2$-Gaussian            & 50.766 & 0.996297 & 46.496 & 0.981445 \\
$X^2$-Gaussian            & 49.264 & 0.996046 & 49.744 & 0.992185 \\
\quad + Cal.              & 50.042 & 0.996662 & 49.744 & 0.992185 \\
\midrule
Ours-Recon                & 49.251 & 0.995681
                          & \textbf{50.331} & \textbf{0.992605} \\
Ours-Full                 & \textbf{52.622} & \textbf{0.997326}
                          & \textbf{50.331} & \textbf{0.992605} \\
\midrule
$\Delta$ Full vs.\ $X^2$  & $+3.359$ & $+0.001280$
                          & $+0.587$ & $+0.000420$ \\
\bottomrule
\end{tabular*}
\end{table}


\begin{table*}[t]
\centering
\small
\renewcommand{\arraystretch}{1.15}
\setlength{\tabcolsep}{5pt}
\caption{Scale-sensitive uncalibrated projection fidelity,
averaged equally across cases.
RMSE, MAE, and relative L2 use the prepared projection scale,
while fixed-reference PSNR uses one reference peak per case.
Gray shading and bold indicate the best result
within each benchmark.}
\label{tab:rawscale}

\begin{tabular*}{\textwidth}{@{\extracolsep{\fill}}llrrrr@{}}
\toprule
Benchmark
& Method
& Raw RMSE $\downarrow$
& Raw MAE $\downarrow$
& \shortstack{Relative L2\\(\%) $\downarrow$}
& \shortstack{Fixed-ref. PSNR\\(dB) $\uparrow$} \\
\midrule
Public
& $X^2$-Gaussian
& 5.543e-04
& 2.841e-04
& 2.261
& 46.243 \\

Public
& Ours-Recon
& \cellcolor[gray]{0.90}\textbf{4.304e-04}
& \cellcolor[gray]{0.90}\textbf{2.326e-04}
& \cellcolor[gray]{0.90}\textbf{1.651}
& \cellcolor[gray]{0.90}\textbf{48.219} \\
\midrule
Private
& $X^2$-Gaussian
& 1.135e-04
& \cellcolor[gray]{0.90}\textbf{4.838e-05}
& 1.155
& 50.995 \\

Private
& Ours-Recon
& \cellcolor[gray]{0.90}\textbf{1.078e-04}
& 4.957e-05
& \cellcolor[gray]{0.90}\textbf{1.098}
& \cellcolor[gray]{0.90}\textbf{51.411} \\
\bottomrule
\end{tabular*}
\end{table*}

\subsection{Experimental Setup}
\label{sec:settings}
\label{sec:benchmarks}

\paragraph{Experimental setup}
We evaluate FlowSlide-GS on thirteen public cases from DIR, 4DLung, and SPARE, and ten clinical 4DCT cases with simulated cone-beam projections. Comparisons include FDK~\cite{feldkamp1984fdk}, NAF~\cite{zha2022naf}, SAX-NeRF~\cite{cai2024saxnerf}, R$^2$-Gaussian~\cite{zha2024r2gaussian}, and $X^2$-Gaussian~\cite{yu2025x2gaussian}. Ours-Recon denotes the reconstructed volumes, while Ours-Full additionally calibrates projections without changing the volumes. We report case-averaged PSNR and SSIM~\cite{wang2004ssim} for held-out projections and reconstructed volumes. Dataset geometry, splits, training schedules, metric definitions, and implementation details are provided in the Appendix.

\subsection{Reconstruction Quality}
\label{sec:reconstruction}
\paragraph{Public and private benchmarks}
\label{sec:public}
\label{sec:private}
Tables~\ref{tab:public} and~\ref{tab:private} report results on thirteen public cases and ten private clinical 4D CT cases with simulated projections. Each benchmark average weights cases equally. Ours-Recon achieves the highest volumetric PSNR and SSIM on all three public subsets and on the private benchmark. Compared with $X^2$-Gaussian, Ours-Recon improves mean volume PSNR from 37.409 to 39.054\,dB on the public benchmark and from 49.744 to 50.331\,dB on the private benchmark, corresponding to gains of 1.645 and 0.587\,dB, respectively. The public-subset gains are 1.475\,dB on DIR, 1.238\,dB on 4DLung, and 2.606\,dB on SPARE. Mean volume SSIM increases from 0.948 to 0.957 on the public benchmark and from 0.992185 to 0.992605 on the private benchmark. These results compare the methods at their respective optimization budgets.

Projection calibration increases mean 2D PSNR from 46.600 to 49.126\,dB on the public benchmark and from 49.251 to 52.622\,dB on the private benchmark, with the reconstructed volumes held fixed. On the public benchmark, Ours-Full exceeds the uncalibrated $X^2$-Gaussian baseline by 2.249\,dB in 2D PSNR. On the private benchmark, Ours-Full exceeds $X^2$-Gaussian by 2.580\,dB when both methods use the same calibration procedure.



\paragraph{Amplitude and structural fidelity}
\label{sec:rawstructure}
The raw-scale evaluation in Table~\ref{tab:rawscale} checks the uncalibrated projections without independent peak normalization. Our model reduces mean squared error by 42.3\% on public data and 10.5\% on private data. On the four-case pilot, all three spatial-gradient measures also improve in every case (in appendix). These measurements complement the volume scores by evaluating projection magnitude and spatial derivatives.

\begin{figure*}[t]
\centering
\includegraphics[width=0.85\textwidth]{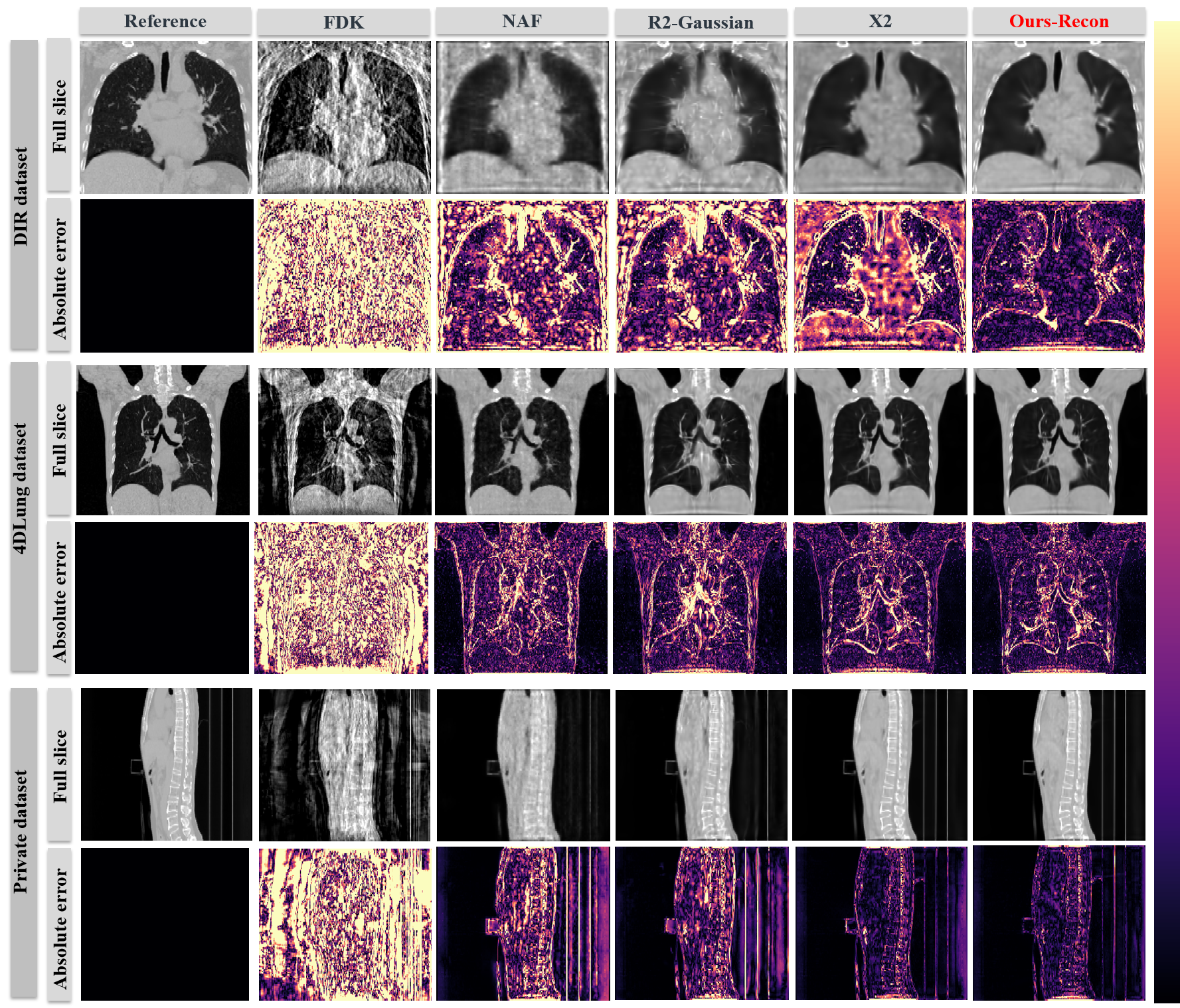}
\caption{Qualitative volume comparisons on the DIR, 4DLung, and private datasets.}
\label{fig:qualitative}
\end{figure*}

\paragraph{Qualitative comparisons}
\label{sec:qualitative}
Figure~\ref{fig:qualitative} shows that FlowSlide-GS generally reduces local reconstruction errors across the DIR, 4DLung, and private datasets, with the clearest improvement near lung boundaries and central interfaces. Improvements are more modest on 4DLung and private cases, and some fine structures remain smoothed. Additional examples and display details are provided in the Appendix.

\begin{table}[t]
\centering
\footnotesize
\setlength{\tabcolsep}{3pt}
\renewcommand{\arraystretch}{1.15}
\caption{Spatial component ablation with 2,000 recovery updates from the common 82k checkpoint. Each cell reports 2D (top) and 3D (bottom) PSNR in dB.}
\label{tab:motionablation}
\begin{tabular*}{\linewidth}{@{\extracolsep{\fill}}lcccc@{}}
\toprule
Case
& \shortstack{All\\components}
& \shortstack{No spatial\\branch}
& \shortstack{Random\\support}
& \shortstack{Tangent\\only}\\
\midrule
DIR 1
& \shortstack{\textbf{44.560}\\\textbf{46.840}}
& \shortstack{44.255\\46.829}
& \shortstack{44.089\\46.782}
& \shortstack{44.558\\46.839}\\
\addlinespace[3pt]
4DLung 1
& \shortstack{\textbf{47.777}\\\textbf{35.956}}
& \shortstack{47.690\\35.947}
& \shortstack{47.676\\35.942}
& \shortstack{47.776\\\textbf{35.956}}\\
\addlinespace[3pt]
SPARE 1
& \shortstack{\textbf{48.213}\\\textbf{33.641}}
& \shortstack{47.856\\33.575}
& \shortstack{47.905\\33.593}
& \shortstack{48.201\\\textbf{33.641}}\\
\addlinespace[3pt]
Private P01
& \shortstack{\textbf{51.049}\\53.261}
& \shortstack{51.005\\53.259}
& \shortstack{51.010\\\textbf{53.263}}
& \shortstack{51.048\\53.262}\\
\midrule
Mean
& \shortstack{\textbf{47.900}\\\textbf{42.425}}
& \shortstack{47.701\\42.403}
& \shortstack{47.670\\42.395}
& \shortstack{47.896\\42.424}\\
\bottomrule
\end{tabular*}
\end{table}

\subsection{Spatial and Temporal Refinement}
\label{sec:ablations}
\paragraph{Controlled recovery protocol}
We study component behavior on four fixed cases: DIR case 1, 4DLung case 1, SPARE case 1, and private P01. Every branch starts from the same trained 82k checkpoint for its case and receives 2,000 additional updates with a fresh optimizer and no densification. The control retaining all reconstruction components receives the same recovery budget. Spatial and auxiliary controls use balanced sampling over all ten phases; the temporal pair uses the same saved hard-phase mask and sampler. Projection scores in these experiments are uncalibrated. This protocol measures removal-and-recovery behavior around a final trained model; the remaining trained components can compensate for a removed branch.

\paragraph{Spatial support and directional correction.}
Table~\ref{tab:motionablation} compares residual-guided support with random
support at the same 15\% capacity, removal of spatial correction, and
removal of the normal component. Relative to random support, residual
guidance improves mean projection PSNR by 0.230\,dB and mean volume
PSNR by 0.030\,dB. Removing the normal component changes volume PSNR
by approximately 0.001\,dB. In the full-benchmark comparison, the
recovered variant without spatial correction differs from the original
full checkpoint by $+0.002$\,dB on the public benchmark and
$-0.018$\,dB on the private benchmark in volume PSNR. The supplementary
material reports the corresponding SSIM and case-level results.

\begin{table}[t]
\centering
\scriptsize
\caption{Full-benchmark checkpoint recovery for the spatial branch.}
\label{tab:spatialall}
\renewcommand{\arraystretch}{1.15}
\setlength{\tabcolsep}{2pt}
\begin{tabular*}{\linewidth}{@{\extracolsep{\fill}}llrrrr@{}}
\toprule
Benchmark & Variant
& \shortstack{2D\\PSNR}
& \shortstack{2D\\SSIM}
& \shortstack{3D\\PSNR}
& \shortstack{3D\\SSIM}\\
\midrule
Public (13) & Ours
& \cellcolor[gray]{0.90}\textbf{46.600}
& 0.994017
& 39.054
& 0.957436\\
Public (13) & No spatial
& 46.534
& \cellcolor[gray]{0.90}\textbf{0.994076}
& \cellcolor[gray]{0.90}\textbf{39.056}
& \cellcolor[gray]{0.90}\textbf{0.957481}\\
\midrule
Private (10) & Ours
& \cellcolor[gray]{0.90}\textbf{49.251}
& \cellcolor[gray]{0.90}\textbf{0.995681}
& \cellcolor[gray]{0.90}\textbf{50.331}
& 0.992605\\
Private (10) & No spatial
& 49.141
& 0.995666
& 50.313
& \cellcolor[gray]{0.90}\textbf{0.992607}\\
\midrule
All (23) & Ours
& \cellcolor[gray]{0.90}\textbf{47.753}
& 0.994741
& \cellcolor[gray]{0.90}\textbf{43.957}
& 0.972727\\
All (23) & No spatial
& 47.668
& \cellcolor[gray]{0.90}\textbf{0.994767}
& 43.951
& \cellcolor[gray]{0.90}\textbf{0.972753}\\
\bottomrule
\end{tabular*}
\end{table}

\paragraph{Full-benchmark spatial recovery}
Across all 23 cases, removing the spatial branch after the 82k checkpoint and applying 2,000 recovery updates reduces 2D PSNR by 0.067\,dB on public data and 0.109\,dB on private data, with negligible 3D changes ($+0.002/-0.018$\,dB). Removing covariance and material adaptation decreases mean 2D/3D PSNR by 0.114/0.020\,dB. These results indicate modest spatial-branch benefits, while the larger 2D gain mainly arises from projection calibration. Detailed results are provided in the appendix.

\begin{table}[t]
\centering
\small
\caption{Effect of phase-isolated refinement after 2,000 recovery updates.
$\Delta$ denotes the change in volume PSNR relative to the common
checkpoint.}
\label{tab:phasecontrols}
\setlength{\tabcolsep}{3pt}
\renewcommand{\arraystretch}{1.12}
\resizebox{\linewidth}{!}{%
\begin{tabular}{@{}lrrrrr@{}}
\toprule
& \multicolumn{2}{c}{PSNR (dB) $\uparrow$}
& \multicolumn{3}{c}{$\Delta$ Volume PSNR (dB) $\uparrow$} \\
\cmidrule(lr){2-3}\cmidrule(lr){4-6}
Update rule & Projection & Volume & Selected
& \shortstack{Inactive\\mean}
& \shortstack{Inactive\\worst} \\
\midrule
Shared parameters
& 47.062
& 42.191
& \cellcolor[gray]{0.90}\textbf{+0.329}
& $-0.365$
& $-1.323$ \\
Phase-isolated (ours)
& \cellcolor[gray]{0.90}\textbf{47.859}
& \cellcolor[gray]{0.90}\textbf{42.418}
& +0.002
& \cellcolor[gray]{0.90}\textbf{+0.000}
& \cellcolor[gray]{0.90}\textbf{+0.000} \\
\bottomrule
\end{tabular}%
}
\end{table}

\begin{table}[t]
\centering
\footnotesize
\caption{Projection calibration on the public and private benchmarks.}
\label{tab:calibrationcontrols}
\renewcommand{\arraystretch}{1.15}
\setlength{\tabcolsep}{2.5pt}
\begin{tabular*}{\linewidth}{@{\extracolsep{\fill}}lrrrr@{}}
\toprule
& \multicolumn{2}{c}{Public (13 cases)}
& \multicolumn{2}{c}{Private (10 cases)}\\
\cmidrule(lr){2-3}\cmidrule(lr){4-5}
Model / calibration
& PSNR $\uparrow$ & SSIM $\uparrow$
& PSNR $\uparrow$ & SSIM $\uparrow$\\
\midrule
$X^2$-Gaussian
& 46.877 & 0.994384 & 49.264 & 0.996046\\
X2 + calibration
& 47.450 & 0.994835 & 50.042 & 0.996662\\
\addlinespace[4pt]
Ours-Recon
& 46.600 & 0.994017 & 49.251 & 0.995681\\
Ours + single gain
& 49.028 & 0.995038 & 52.504 & 0.997294\\
Ours + coarse/detail
& 49.126
& \cellcolor[gray]{0.90}\textbf{0.995068}
& 52.622 & 0.997326\\
Ours + no LOO
& \cellcolor[gray]{0.90}\textbf{49.154}
& 0.994980
& \cellcolor[gray]{0.90}\textbf{52.854}
& \cellcolor[gray]{0.90}\textbf{0.997393}\\
Ours + nested validation
& 49.126
& \cellcolor[gray]{0.90}\textbf{0.995068}
& 52.623 & 0.997326\\
\bottomrule
\end{tabular*}
\end{table}

\begin{figure}{} 
\centering
	\includegraphics[scale=0.3]{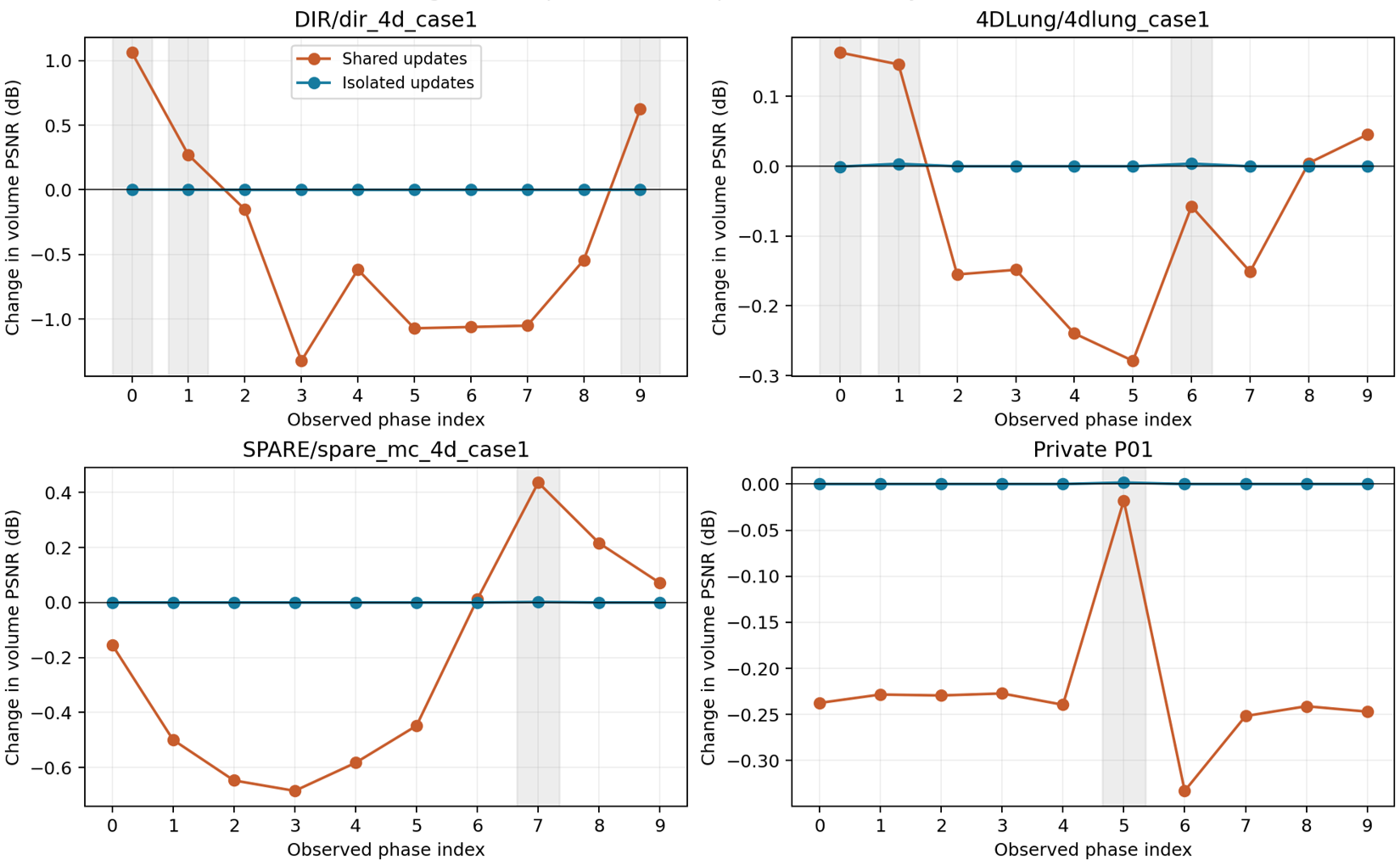}
	\caption{Phase-wise volume-PSNR changes after the seed-0 temporal recovery comparison, relative to the common 82k checkpoint.}
	\label{fig:phaseupdateeffects}
\end{figure}

\paragraph{Phase-isolated updates}
Shared refinement updates the mature shared model with the phase dictionary frozen. Phase-isolated refinement freezes the shared model and updates the dictionary. Table~\ref{tab:phasecontrols} and Fig.~\ref{fig:phaseupdateeffects} show that shared updates improve selected-phase volume PSNR by 0.329\,dB and reduce inactive-phase PSNR by 0.365\,dB on average, with a worst decrease of 1.323\,dB. Phase-isolated updates preserve all 32 inactive case--phase pairs and improve the selected phases by 0.002\,dB. Their overall advantage in this experiment therefore comes primarily from preserving the inactive phases. The supplementary material reports recovery results for three seeds using the same trained checkpoints.

\subsection{Projection Calibration}
\label{sec:calibrationresults}
Calibration improves normalized projection PSNR by 2.526\,dB on public data and 3.372\,dB on private data, with the reconstruction fixed. To test whether the benefit is specific to our backbone, we apply the identical module and settings to X2. Table~\ref{tab:calibrationcontrols} shows that both backbones benefit, while our complete model remains 1.676\,dB higher on public data and 2.580\,dB higher on private data. This comparison separates the benefit of adding a corrector from the difference between the resulting pipelines.

\begin{figure}{} 
\centering
	\includegraphics[scale=0.36]{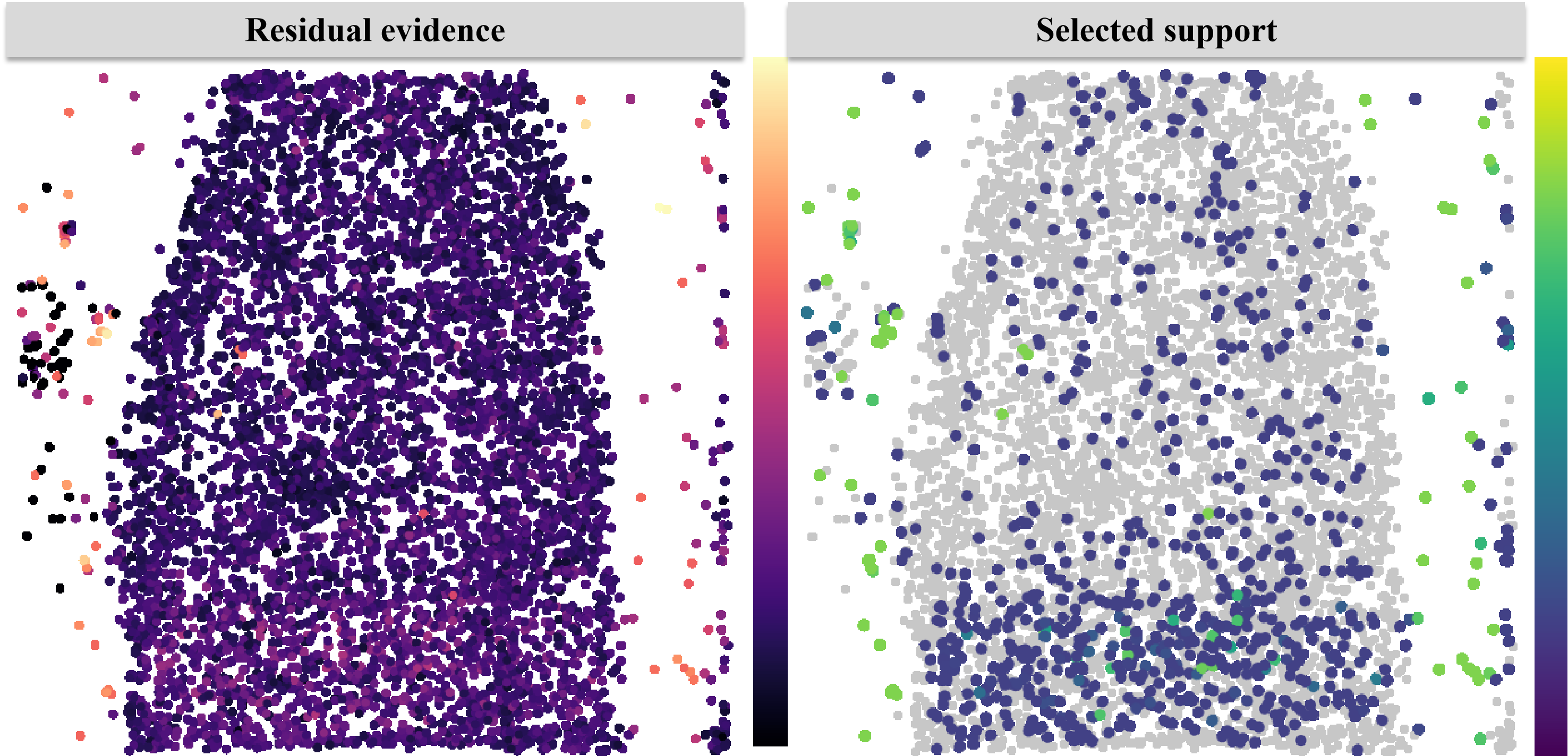}
	\caption{Residual evidence and selected Gaussian primitives for local
refinement. Both panels show the same sampled primitives in a canonical
slab. Left: normalized projection-residual evidence accumulated as an
exponential moving average. Right: selected primitives are colored by
their support gate values, with unselected primitives shown in gray.}
\label{fig:mechanismevidence}
\vspace{-3mm}
\end{figure}

\paragraph{Spatial allocation of local refinement.}
Figure~\ref{fig:mechanismevidence} shows how the selector allocates
local refinement across Gaussian primitives. In this example, selected
primitives are concentrated toward the lower part of the displayed slab,
with additional selections scattered throughout the slab. In the four-case
recovery experiment, residual-guided selection improves projection PSNR
by 0.230\,dB and volume PSNR by 0.030\,dB over random selection at the
same 15\% support capacity. The observed benefit is therefore more
pronounced in projection fidelity.

\paragraph{Calibration design}
Separate coarse/detail gains improve PSNR over a tied gain by 0.098\,dB on public data and 0.119\,dB on private data. Removing LOO exclusion gives the highest PSNR but lowers public SSIM, indicating that LOO prevents self-residual features during gain fitting without improving test PSNR in this experiment. A nested training--validation control produces nearly identical results. Qualitative comparisons, anchor perturbations, and implementation details are provided in the appendix.

\subsection{Limitations}
\label{sec:limitations}
Our evaluation is limited to prepared or simulated projections and does not cover real acquisition artifacts. Independent evaluation and repeated training are needed to assess generalization. Moreover, projection calibration does not enforce volumetric consistency, and periodic closure does not guarantee temporal smoothness or global invertibility. Further analyses are provided in the Appendix.

\section{Conclusion}
\label{sec:conclusion}
We introduced FlowSlide-GS, which combines periodic Gaussian transport, residual-guided spatial adaptation, phase-isolated refinement, and projection correction for 4D CT reconstruction. The method improves volumetric fidelity, while calibration enhances projection quality and phase isolation prevents cross-phase degradation. Future work will address computational cost, temporal smoothness, and evaluation on acquired projection data.

\section{Use of Large Language Models}
During manuscript preparation, large language models (LLMs) were used to assist with translation, language editing, and \LaTeX{} syntax. LLMs also assisted in implementing selected method modules and visualization tools. All AI-assisted content was reviewed and verified by the authors, who take full responsibility for the manuscript, including AI-assisted code and figures.

\setcounter{page}{1}
\renewcommand{\thepage}{S\arabic{page}}
\setcounter{section}{0}
\setcounter{subsection}{0}
\setcounter{figure}{0}
\setcounter{table}{0}
\setcounter{equation}{0}
\renewcommand{\thesection}{\Alph{section}}
\renewcommand{\thefigure}{S\arabic{figure}}
\renewcommand{\thetable}{S\arabic{table}}
\renewcommand{\theequation}{S\arabic{equation}}
\ifdefined\theHsection
  \renewcommand{\theHsection}{supp.\Alph{section}}
  \renewcommand{\theHfigure}{supp.\arabic{figure}}
  \renewcommand{\theHtable}{supp.\arabic{table}}
  \renewcommand{\theHequation}{supp.\arabic{equation}}
\fi
\maketitlesupplementary

This supplement provides the acquisition and evaluation protocols (Sec.~\ref{app:protocol}), details of phase isolation and projection calibration (Sec.~\ref{app:algorithm}), additional controlled experiments (Sec.~\ref{app:additionalrecovery}), reconstruction and motion diagnostics (Sec.~\ref{app:diagnostics}), computational measurements (Sec.~\ref{app:runtime}), case-level results (Sec.~\ref{app:cases}), and qualitative comparisons (Sec.~\ref{app:qualitative}). Throughout, Ours-Recon and Ours-Full share exactly the same reconstructed volumes; the latter additionally calibrates normalized projections.

\section{Experimental Protocol and Implementation}
\label{app:protocol}

\subsection{Datasets, Geometry, and Held-Out Views}
The public benchmark contains five DIR, five 4DLung, and three SPARE cases from the prepared $X^2$-Gaussian release~\cite{yu2025x2gaussian}. DIR originates from the respiratory registration resource of Castillo et al.~\cite{castillo2009landmarks}. We retain the released camera geometry, train/test partition, native reference grids, and Gaussian initialization. Each case has ten reference phases, 300 training projections, and 100 held-out projections. Table~\ref{supp:tab:data} summarizes the benchmarks, and Table~\ref{supp:tab:publicgeometry} lists the native public grids.

\begin{table*}[h]
\centering\small
\caption{Benchmark configuration. Train/test counts are per case.
Private projections are simulated from reconstructed clinical CT phases.}
\label{supp:tab:data}
\setlength{\tabcolsep}{4pt}
\begin{tabular*}{\textwidth}{@{\extracolsep{\fill}}llrrll@{}}
\toprule
Benchmark & Subset & Cases & Train/test & Volume grid & Detector\\
\midrule
Public & DIR & 5 & 300/100 & $256\times256\times(94\text{--}112)$ & $512\times512$\\
Public & 4DLung & 5 & 300/100 & $512\times512\times(91\text{--}135)$ & $1024\times1024$\\
Public & SPARE & 3 & 300/100 & Case-specific & $512\times512$\\
Private & Clinical 4D CT & 10 & 300/100 & $256\times256\times128$ & $512\times512$\\
\bottomrule
\end{tabular*}
\end{table*}

\paragraph{Private data preparation.}
The private source comprises ten respiratory-resolved clinical CT studies,
denoted P01--P10, with reconstructed phases at 0\%, 10\%, \ldots, 90\%
of the respiratory cycle. Averaged CT, maximum-intensity projections,
planning CT, and radiotherapy objects are excluded. Phase grids are checked
for consistency and resampled to $256\times256\times128$.
We simulate 400 sequential cone-beam views using a $512\times512$ detector,
holding out one view in four within each phase. This gives 30 training
and ten test views per phase. The starting time of each phase bin defines
its representative timestamp. All methods receive the same converted
measurements; $X^2$-Gaussian and ours also share the prepared Gaussian
initialization. The experiment therefore evaluates simulated measurements
of clinical anatomy.

\paragraph{Public geometry.}
All public headers specify source--detector and source--origin distances
of 1500 and 1000, respectively, in the millimeter scanner convention,
with $1\times1$ detector sampling. Public and private angle-header
conventions differ; the actual camera angles define the acquisition
geometry. Stored array axes are retained during evaluation.

\begin{table}[h]
\centering\small
\caption{Native public reconstruction grids and voxel spacing. Dimensions follow the stored axis ordering.}
\label{supp:tab:publicgeometry}
\begin{tabular}{llrr}
\toprule
Subset & Case & Volume dimensions & Voxel spacing (mm)\\
\midrule
DIR & 1 & $256\times256\times94$ & $(0.97,0.97,2.5)$\\
DIR & 2 & $256\times256\times112$ & $(1.16,1.16,2.5)$\\
DIR & 3 & $256\times256\times104$ & $(1.15,1.15,2.5)$\\
DIR & 4 & $256\times256\times99$ & $(1.13,1.13,2.5)$\\
DIR & 5 & $256\times256\times106$ & $(1.10,1.10,2.5)$\\
4DLung & 1 & $512\times512\times135$ & $(0.9766,0.9766,3.0)$\\
4DLung & 2 & $512\times512\times104$ & $(0.9766,0.9766,3.0)$\\
4DLung & 3 & $512\times512\times94$ & $(0.9766,0.9766,3.0)$\\
4DLung & 4 & $512\times512\times91$ & $(0.9766,0.9766,3.0)$\\
4DLung & 5 & $512\times512\times126$ & $(0.9766,0.9766,3.0)$\\
SPARE & 1 & $281\times319\times220$ & $(1.0,1.0,1.0)$\\
SPARE & 2 & $208\times288\times220$ & $(1.0,1.0,1.0)$\\
SPARE & 3 & $208\times288\times220$ & $(1.0,1.0,1.0)$\\
\bottomrule
\end{tabular}
\end{table}

\subsection{Metrics and Aggregation}
\label{supp:sec:metrics}
Reconstructions are evaluated at the reference phase centers and on the native evaluation grids, without post-hoc registration. Reference CT volumes and held-out projections are used only for evaluation.

\paragraph{Normalized projection metrics.}
Following the $X^2$-Gaussian evaluator, predictions and references are
independently peak-normalized. Let
\begin{equation}
N(y)=\frac{y}{\max(\max_u y(u),\epsilon)}.
\label{supp:eq:normalization}
\end{equation}
Projection PSNR is computed with a unit peak,
\begin{equation}
\operatorname{PSNR}_{2D}
=-10\log_{10}\operatorname{MSE}\!\left(N(\widehat y),N(y)\right).
\end{equation}
SSIM~\cite{wang2004ssim} uses the corresponding normalized images. Ours-Full applies the saved calibrator to the prediction before projection evaluation. Independent normalization emphasizes projection shape; scale-sensitive measurements are defined in Sec.~\ref{app:diagnostics}.

\paragraph{Volume metrics.}
Volume PSNR uses the prepared attenuation scale with a unit peak. For volume SSIM, nonempty reference slices are averaged within each of the three orthogonal axes, then the three axis means are averaged. Projection metrics are averaged over held-out views, and volume metrics over the ten reference phases. Benchmark means give every case equal weight. In particular, the public mean weights DIR, 4DLung, and SPARE by $5/13$, $5/13$, and $3/13$, rather than weighting the three subsets equally.

\paragraph{Frozen-volume identity.}
Ours-Full reuses the Ours-Recon Gaussian checkpoint and its 3D evaluation results; only its projection predictions are recalculated. Their matching volume scores express an identity of the evaluated representation. Calibration does not provide an additional volume-quality gain. Best values in the supplementary tables are marked separately for each metric; ties at the displayed precision are marked together.

\subsection{Architecture and Staged Optimization}
Each case is optimized independently with Adam~\cite{kingma2015adam}. The reconstruction backbone uses two-layer coordinate networks with 48 hidden channels, three temporal harmonics, and low/high spatial frequency counts of three/six. The final midpoint integrator uses two steps. The spatial branch has 24 hidden channels and selects 15\% of the primitives. The phase-local branch has 16 hidden channels, ten phase bins, and raised-cosine interpolation.

Reconstruction runs for 82,000 updates. At iteration 76,000, the shared reconstruction model and spatial adapter are frozen, and the final 6,000 updates optimize phase-local capacity. The projection calibrator is fitted after reconstruction with the complete 3D model fixed. The main $X^2$-Gaussian and proposed reconstructions initialize Python, NumPy, and PyTorch random seeds to zero. Seeds 17 and 42 are additionally used for the recovery controls in Sec.~\ref{supp:sec:temporal}.

\subsection{Baseline Implementations and Budgets}
FDK~\cite{feldkamp1984fdk} uses TIGRE~\cite{biguri2016tigre} with a Ram--Lak filter. Static R$^2$-Gaussian~\cite{zha2024r2gaussian} initializes 50,000 points from a training-only FDK reconstruction within each phase. NAF~\cite{zha2022naf} and SAX-NeRF~\cite{cai2024saxnerf} retain their official networks, encoders, ray sampling, and rendering implementations. Dataset adapters set geometry-derived network bounds and expose the common floating-point evaluation protocol. Each static method reconstructs ten independent phase volumes. Table~\ref{supp:tab:baselineprotocol} summarizes the executed schedules.

\begin{table*}[h!]
\centering\small
\caption{Baseline representations and optimization schedules.}
\label{supp:tab:baselineprotocol}
\renewcommand{\arraystretch}{1.08}
\setlength{\tabcolsep}{5pt}
\begin{tabular*}{\textwidth}{@{\extracolsep{\fill}\hspace{\tabcolsep}}lll}
\toprule
Method & Representation & Training schedule\\
\midrule
\multicolumn{3}{l}{\textit{Independent phase models}}\\[2pt]
FDK & Per-phase volume & Analytic reconstruction\\
R$^2$-Gaussian & Per-phase Gaussians & 30,000 updates/phase\\
NAF & Per-phase attenuation field & $3000n_p$ updates/phase\\
SAX-NeRF & Per-phase Lineformer & $1500n_p$ updates/phase\\
\midrule
\multicolumn{3}{l}{\textit{Shared dynamic models}}\\[2pt]
$X^2$-Gaussian & Dynamic Gaussians & 30,000 total iterations\\
Ours-Recon & Shared and phase-local motion & 82,000 total iterations\\
Ours-Full & Same volume; projection calibration & 82,000 iterations + calibration fit\\
\bottomrule
\end{tabular*}
\end{table*}

With a one-view batch, NAF and SAX-NeRF perform $3000n_p$ and $1500n_p$ updates per phase, where $n_p$ is the number of training views in phase $p$. The private bins contain 30 training views, giving 90,000 and 45,000 updates per phase; public bin sizes vary. These schedules differ from the 150,000 NeRF iterations reported in the $X^2$-Gaussian supplementary comparison~\cite{yu2025x2gaussian}, so the static results describe the local configurations stated here.

The $X^2$-Gaussian changes for NumPy pickles, checkpoint resumption, CPU accumulation of evaluation images, and removal of an initial evaluation call. Its reconstruction model and loss definitions are unchanged. Its mean volume PSNR on DIR, 4DLung, and SPARE is 39.288, 38.606, and 32.284\,dB, respectively,
compared with the reported 39.34, 38.61, and 32.24\,dB. Optimization budgets differ across methods; the separate equal-time experiment is reported in Sec.~\ref{supp:sec:equaltime}.

\begin{table*}[h]
\centering\small
\caption{Auxiliary-module removal with 2,000 recovery updates from the common 82k checkpoint on four cases. Means weight cases equally; projection scores are uncalibrated and PSNR is in dB.}
\label{supp:tab:auxiliarycontrols}
\renewcommand{\arraystretch}{1.08}
\setlength{\tabcolsep}{5pt}
\begin{tabular*}{\textwidth}{@{\extracolsep{\fill}\hspace{\tabcolsep}}lrrrr}
\toprule
Configuration & 2D PSNR & 2D SSIM & 3D PSNR & 3D SSIM\\
\midrule
All components & \cellcolor[gray]{0.90}\textbf{47.900} & 0.994628 & \cellcolor[gray]{0.90}\textbf{42.425} & \cellcolor[gray]{0.90}\textbf{0.962878}\\
No covariance adaptation & 47.865 & \cellcolor[gray]{0.90}\textbf{0.994632} & 42.423 & 0.962876\\
No material adaptation & 47.824 & 0.994620 & 42.407 & 0.962854\\
Neither auxiliary adapter & 47.786 & 0.994623 & 42.405 & 0.962849\\
\bottomrule
\end{tabular*}
\end{table*}

\section{Method Details}
\label{app:algorithm}

\subsection{Phase Interpolation and Inactive-Center Preservation}
Let the ten uniformly spaced phase centers be $\tau_p=p/P$, with $P=10$.
Between adjacent centers $\tau_p$ and $\tau_{p+1}$, define the local
circular coordinate $u\in[0,1]$ and interpolation weight
\begin{equation}
\alpha(u)=\tfrac12[1-\cos(\pi u)].
\end{equation}
The nonzero dictionary weights on that interval are
$K_p=1-\alpha(u)$ and $K_{p+1}=\alpha(u)$, with indices taken modulo $P$.
The interval joining the last center to phase one wraps to phase zero.
At any phase center, $K_p(\tau_q)=\delta_{pq}$.

Write the masked residual as
\begin{equation}
\Delta^{\mathrm{phase}}_\tau(x)
=g(x,\tau)b_{\max}\tanh\!\left[
\sum_{p=0}^{P-1}h_pK_p(\tau)c_p(x)\right],
\label{supp:eq:phase_residual}
\end{equation}
where the binary mask $h_p$ selects the active phase coefficients and
$c_p(x)$ are outputs of the shared spatial network.
At an inactive center $\tau_q$,
\begin{equation}
\sum_p h_pK_p(\tau_q)c_p(x)=h_qc_q(x)=0.
\end{equation}
Consequently, $\Delta^{\mathrm{phase}}_{\tau_q}(x)=0$ for every $x$. Because this is identically zero as a function of space, its spatial derivative is also zero wherever the residual is differentiable:
\begin{equation}
\nabla_x\Delta^{\mathrm{phase}}_{\tau_q}(x)=0.
\end{equation}
Thus the added phase map is the identity, with identity spatial Jacobian, at inactive centers. With the shared model, spatial adapter, and mask fixed, training the shared coefficient network cannot change the reconstructed volumes at those centers.

\subsection{Projection Calibration Procedure}
\label{app:calibration}
For each training view, the frozen reconstruction yields a prediction
$\widehat y_j$. Its residual anchor is
$r_j=N(y_j)-N(\widehat y_j)$, using
Eq.~\eqref{supp:eq:normalization}. Anchors are stored as
$128\times128$ normalized residual maps.

\begin{table*}[h]
\centering\small
\caption{Per-case recovery results: uncalibrated projection / volume PSNR (dB). The first three variants use balanced all-phase recovery; the last two use identical saved hard-phase sampling. Bold compares these groups of three and two separately, for each case and metric.}
\label{supp:tab:additionalpercase}
\renewcommand{\arraystretch}{1.08}
\setlength{\tabcolsep}{5pt}
\begin{tabular*}{\textwidth}{@{\extracolsep{\fill}\hspace{\tabcolsep}}lccccc}
\toprule
Case & No covariance & No material & Neither auxiliary & Phase isolated & Phase shared\\
\midrule
DIR case 1 & \textbf{44.524} / \textbf{46.841} & 44.309 / 46.776 & 44.259 / 46.773 & \textbf{44.521} / \textbf{46.835} & 43.254 / 46.449\\
4DLung case 1 & 47.748 / 35.950 & \textbf{47.772} / \textbf{35.956} & 47.745 / 35.950 & \textbf{47.720} / \textbf{35.948} & 47.634 / 35.880\\
SPARE case 1 & 48.143 / 33.635 & \textbf{48.169} / \textbf{33.636} & 48.100 / 33.630 & \textbf{48.163} / \textbf{33.628} & 47.359 / 33.399\\
Private P01 & 51.044 / \textbf{53.265} & \textbf{51.047} / 53.262 & 51.040 / \textbf{53.265} & \textbf{51.030} / \textbf{53.260} & 50.002 / 53.035\\
\bottomrule
\end{tabular*}
\end{table*}

For a query view, the calibrator selects the left and right angular
neighbors within the same phase. The residual is interpolated using their
circular angular distances, as specified in the main paper. A $5\times5$
averaging kernel separates coarse and detail components. During gain
fitting, leave-one-out (LOO) interpolation removes the target training
anchor from its own feature construction. Two phase-specific gains are
fitted with ridge coefficient $10^{-4}$ and clipped to $[0,1.25]$.
At inference, the weighted residual components are bilinearly upsampled
and added to the normalized prediction, followed by clipping to $[0,1]$.
No correction is applied when same-phase angular support is unavailable.

All anchors, fitting targets, and gain-selection data come from training projections. Test projections are reserved for final evaluation. The single-gain control ties the coarse and detail gains; the no-LOO control retains each target anchor when constructing its fitting features. The same calibrator settings are used for the $X^2$-Gaussian control.

\paragraph{Nested training-only validation.}
A fixed phase-stratified fifth of the training views is reserved from both interpolation anchors and gain selection. Inner LOO estimates select the ridge coefficient from $\{10^{-5},10^{-4},10^{-3}\}$, and the outer residual error is evaluated once. Final gains are refitted using all training anchors with the selected coefficient. This procedure changes only calibration; the reconstruction remains fixed.

\section{Additional Controlled Experiments}
\label{app:additionalrecovery}

\subsection{Recovery Protocol and Interpretation}
The four fixed pilot cases are DIR case 1, 4DLung case 1, SPARE case 1, and private P01. Each branch starts from the same trained 82k checkpoint for its case, uses a fresh optimizer, and receives 2,000 additional updates without densification. The control retaining all reconstruction components receives the same recovery budget. Spatial and auxiliary controls use balanced sampling over all ten phases. The temporal pair uses the same saved hard-phase mask and sampler.

\subsection{Spatial Support and Auxiliary Adaptation}
At the same 15\% support capacity, residual-guided support exceeds random support by 0.230\,dB in mean projection PSNR and 0.030\,dB in volume PSNR on the four pilot cases, as reported in the main paper. The effect of removing the normal component is much smaller. The main-paper full-benchmark recovery comparison likewise shows small volume changes: $+0.002$\,dB on public data and $-0.018$\,dB on private data after removing the spatial branch and applying recovery updates.

Table~\ref{supp:tab:auxiliarycontrols} adds covariance and material adaptation controls, including SSIM. Removing covariance or material adaptation decreases mean projection PSNR by 0.035 and 0.076\,dB, respectively, with smaller volume changes. Removing both reduces the reported mean projection and volume PSNR by 0.114 and 0.020\,dB. Some removals slightly improve the private volume score, indicating case-dependent contributions. Table~\ref{supp:tab:additionalpercase} lists the corresponding case measurements.

\subsection{Temporal Recovery Across Seeds}
\label{supp:sec:temporal}
Shared refinement updates the mature shared model with the phase
dictionary frozen. Phase-isolated refinement freezes the shared model
and updates the dictionary. The four starting checkpoints, phase masks,
objective, and 2,000-update budget are fixed across seeds.
Table~\ref{supp:tab:seedrecovery} reports the isolated-minus-shared PSNR
advantages for seeds 0, 17, and 42. Standard deviations describe recovery
stability across these fixed-checkpoint trials.

\begin{table}[t]
\centering\footnotesize
\caption{Three-seed temporal recovery. $\Delta$: isolated minus
shared PSNR (dB), averaged over four cases.}
\label{supp:tab:seedrecovery}
\renewcommand{\arraystretch}{1.08}
\setlength{\tabcolsep}{3pt}
\begin{tabular*}{\linewidth}{@{\extracolsep{\fill}}lrr@{}}
\toprule
Seed & $\Delta$2D PSNR & $\Delta$3D PSNR\\
\midrule
0  & +0.7965 & +0.2269\\
17 & +0.7932 & +0.2265\\
42 & +0.7928 & +0.2254\\
\midrule
Mean $\pm$ SD
& $+0.7942 \pm 0.0020$
& $+0.2263 \pm 0.0007$\\
\bottomrule
\end{tabular*}
\end{table}

The seed-0 comparison in the main paper shows that shared updates improve
selected-phase volume PSNR by 0.329\,dB, while decreasing inactive-phase
PSNR by 0.365\,dB on average, with a worst decrease of 1.323\,dB.
Phase-isolated updates preserve all 32 inactive case--phase pairs and
improve the selected phases by 0.002\,dB. The aggregate advantage of
phase isolation in this experiment therefore comes primarily from
preserving inactive phases, consistent with the center-wise identity
in Sec.~\ref{app:algorithm}.

\subsection{Calibrator-Anchor Sensitivity}
We perturb the calibrator anchors on the four pilot cases while keeping
both reconstruction backbones fixed at their full-data solutions.
The four perturbations are: retaining alternating sorted-angle anchors
within each phase; removing anchors in $[0,60^\circ)$; shifting anchor
phase labels by one bin; and adding Gaussian noise with standard
deviation 0.01 to normalized anchor residuals. Each calibrator is refitted
using its modified anchors and evaluated on unchanged held-out views. Compared with the original anchors, halving their number reduces our
mean projection PSNR from 50.176 to 48.832\,dB. Shifting phase labels
reduces it to 47.915\,dB, slightly below the $X^2$-Gaussian control at
47.959\,dB. Under residual noise, our PSNR remains higher but SSIM is
lower than the control.

\begin{table}[t]
\centering\footnotesize
\caption{Anchor sensitivity with frozen backbones
(four-case means; PSNR in dB).}
\label{supp:tab:anchorstress}
\renewcommand{\arraystretch}{1.08}
\setlength{\tabcolsep}{2.5pt}
\begin{tabular*}{\linewidth}{@{\extracolsep{\fill}}lrrrr@{}}
\toprule
& \multicolumn{2}{c}{Ours}
& \multicolumn{2}{c}{$X^2$}\\
\cmidrule(lr){2-3}\cmidrule(lr){4-5}
Anchor condition & PSNR & SSIM & PSNR & SSIM\\
\midrule
Original
& \cellcolor[gray]{0.90}\textbf{50.18}
& \cellcolor[gray]{0.90}\textbf{0.9957}
& 48.72 & 0.9955\\
Half per phase
& \cellcolor[gray]{0.90}\textbf{48.83}
& \cellcolor[gray]{0.90}\textbf{0.9952}
& 48.28 & 0.9951\\
Remove $[0,60^\circ)$
& \cellcolor[gray]{0.90}\textbf{49.73}
& \cellcolor[gray]{0.90}\textbf{0.9955}
& 48.57 & 0.9953\\
Shift one bin
& 47.92
& \cellcolor[gray]{0.90}\textbf{0.9951}
& \cellcolor[gray]{0.90}\textbf{47.96}
& \cellcolor[gray]{0.90}\textbf{0.9951}\\
Noise $\sigma=0.01$
& \cellcolor[gray]{0.90}\textbf{49.45}
& 0.9944
& 48.41
& \cellcolor[gray]{0.90}\textbf{0.9948}\\
\bottomrule
\end{tabular*}
\end{table}

\section{Reconstruction and Motion Diagnostics}
\label{app:diagnostics}

\subsection{Scale-Sensitive Projection Fidelity}
The main-paper raw-scale table evaluates uncalibrated projections in
prepared-data line-integral units. Both prediction and reference are
divided by the known scene-scale factor, without fitting a gain.
For all held-out pixels within one case, let
\begin{equation}
E_{\mathrm{rel}}=
\left(\frac{\sum_j\|\widehat y_j-y_j\|_2^2}
{\sum_j\|y_j\|_2^2+\epsilon}\right)^{1/2}.
\end{equation}
Raw RMSE and MAE measure absolute errors in those units. Fixed-reference
PSNR uses the maximum reference value across all test views of the case,
shared by both methods. Each metric is calculated per case before
averaging cases equally.

Fixed-reference PSNR increases from 46.243 to 48.219\,dB on public data
and from 50.995 to 51.411\,dB on private data. Private MAE increases
from $4.838\times10^{-5}$ to $4.957\times10^{-5}$ despite lower RMSE and
relative L2. Thus the raw-scale advantage depends on the error measure.

\begin{figure*}[h]
\centering
\includegraphics[width=1.0 \textwidth]{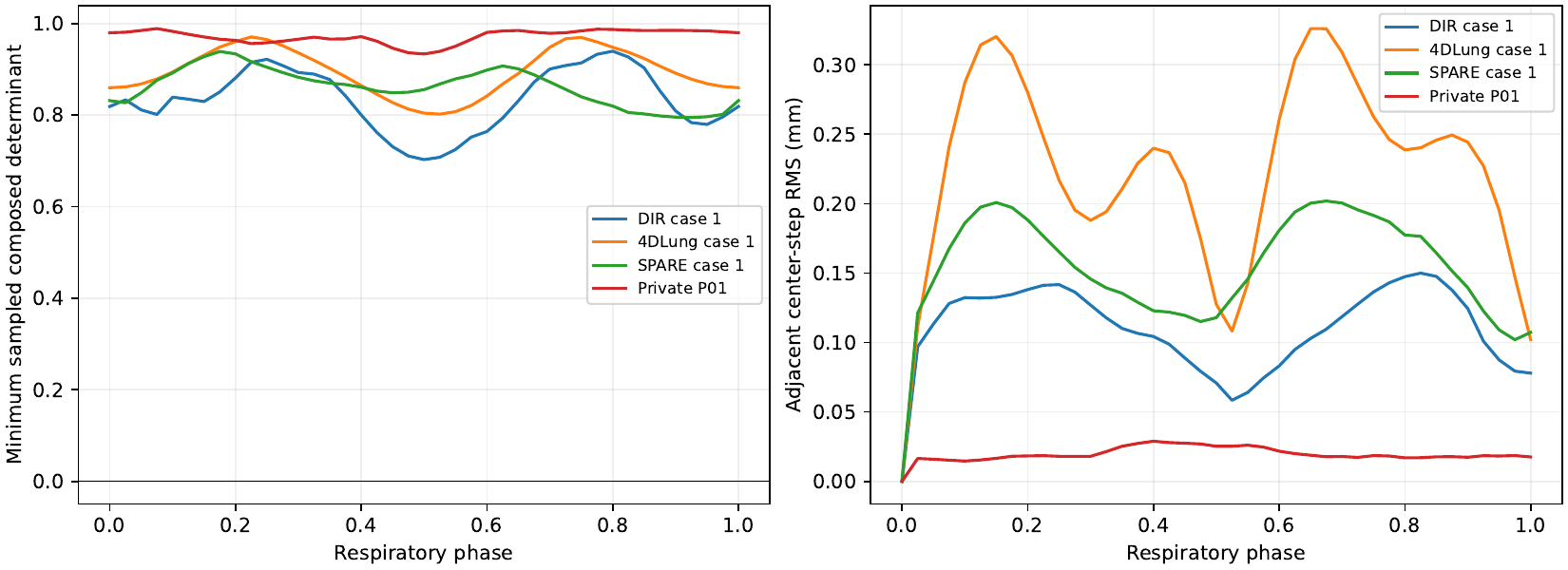}
\caption{Motion diagnostics over 41 respiratory phases for the four pilot cases. Left: minimum composed Jacobian determinant at sampled Gaussian centers. Right: RMS center displacement between adjacent queried phases.}
\label{supp:fig:continuouschecks}
\end{figure*}

\subsection{Spatial-Derivative Fidelity}
We compute central differences using physical voxel spacing over the
common volume interior. Gradient relative L2 compares the full gradient
vectors, gradient cosine is their normalized inner product, and magnitude
relative L1 compares gradient magnitudes. No smoothing, intensity
normalization, edge threshold, or region selection is applied.
Table~\ref{supp:tab:volumegradients} first averages the ten reference
phases within each pilot case, then weights the four cases equally.
All three measures improve in every pilot case. These are derivative
measurements rather than segmentation or motion-ground-truth metrics.

\begin{table*}[h]
\centering\small
\caption{Spatial-derivative fidelity over all ten observed phases of four cases. Central differences use physical voxel spacing over the common volume interior. Metrics average phases within each case, then weight cases equally.}
\label{supp:tab:volumegradients}
\renewcommand{\arraystretch}{1.08}
\setlength{\tabcolsep}{5pt}
\begin{tabular*}{\textwidth}{@{\extracolsep{\fill}\hspace{\tabcolsep}}llrrr}
\toprule
Case & Method & Gradient rel. L2 $\downarrow$ & Gradient cosine $\uparrow$ & Magnitude rel. L1 $\downarrow$\\
\midrule
DIR case 1 & $X^2$-Gaussian & 0.5424 & 0.8423 & 0.4733\\
DIR case 1 & Ours-Recon & \cellcolor[gray]{0.90}\textbf{0.3974} & \cellcolor[gray]{0.90}\textbf{0.9189} & \cellcolor[gray]{0.90}\textbf{0.3643}\\
4DLung case 1 & $X^2$-Gaussian & 0.4719 & 0.8832 & 0.4896\\
4DLung case 1 & Ours-Recon & \cellcolor[gray]{0.90}\textbf{0.3992} & \cellcolor[gray]{0.90}\textbf{0.9180} & \cellcolor[gray]{0.90}\textbf{0.4305}\\
SPARE case 1 & $X^2$-Gaussian & 0.5152 & 0.8581 & 0.3946\\
SPARE case 1 & Ours-Recon & \cellcolor[gray]{0.90}\textbf{0.4079} & \cellcolor[gray]{0.90}\textbf{0.9133} & \cellcolor[gray]{0.90}\textbf{0.3324}\\
Private P01 & $X^2$-Gaussian & 0.2624 & 0.9654 & 0.2511\\
Private P01 & Ours-Recon & \cellcolor[gray]{0.90}\textbf{0.2213} & \cellcolor[gray]{0.90}\textbf{0.9752} & \cellcolor[gray]{0.90}\textbf{0.2245}\\
Four-case mean & $X^2$-Gaussian & 0.4480 & 0.8872 & 0.4022\\
Four-case mean & Ours-Recon & \cellcolor[gray]{0.90}\textbf{0.3565} & \cellcolor[gray]{0.90}\textbf{0.9314} & \cellcolor[gray]{0.90}\textbf{0.3379}\\
\bottomrule
\end{tabular*}
\end{table*}

\subsection{Continuous-Phase and Cycle Checks}
\label{app:motion}
We query the final model at 41 fixed phases over one cycle, including
intermediate phases. Table~\ref{supp:tab:cycleconsistency} compares
centers, packed covariances, and density at phases zero and one.
It also reports the minimum composed Jacobian determinant over the
sampled Gaussian centers and phases. Endpoint discrepancies are
numerically zero, and no nonpositive determinant is sampled.
Figure~\ref{supp:fig:continuouschecks} shows the determinant and
adjacent-center-step traces.

\begin{table}[t]
\centering\footnotesize
\caption{Cycle diagnostics over 41 phases.
Endpoint errors compare phases 0 and 1; seam RMS measures
derivative mismatch (mm/cycle).}
\label{supp:tab:cycleconsistency}
\renewcommand{\arraystretch}{1.08}
\setlength{\tabcolsep}{2pt}
\begin{tabular*}{\linewidth}{@{\extracolsep{\fill}}lrrrrr@{}}
\toprule
Case
& \shortstack{Center\\RMS (mm)}
& \shortstack{Cov.\\rel. L2}
& \shortstack{Density\\rel. L2}
& \shortstack{Min.\\$\det J$}
& \shortstack{Seam\\RMS}\\
\midrule
DIR 1    & 0 & 0 & 0 & 0.7026 & 52.81\\
4DLung 1 & 0 & 0 & 0 & 0.8019 & 40.41\\
SPARE 1  & 0 & 0 & 0 & 0.7946 & 80.35\\
P01      & 0 & 0 & 0 & 0.9333 & 9.640\\
\bottomrule
\end{tabular*}
\end{table}

The seam diagnostic compares one-sided finite-difference center
derivatives using phase step $10^{-4}$. Its nonzero values distinguish
endpoint closure from temporal derivative continuity. Positive sampled
determinants do not establish global invertibility. Moreover, reference
volumes are available only at the ten observed phase centers, so the
intermediate-phase queries assess numerical behavior rather than
intermediate-phase reconstruction accuracy.

\section{Computational Cost}
\label{app:runtime}

\subsection{Inference Latency and Memory}
Table~\ref{supp:tab:inferencecontrols} compares both methods on the same NVIDIA RTX 6000 Ada GPU over the four pilot cases. Each operation follows two preparation calls and uses CUDA synchronization. We time ten first-held-out-view renders and three native-grid phase-zero volume queries per case, then average the case medians. Rendering includes output transfer to CPU; queried volumes remain on the GPU.

For this timing experiment, $X^2$-Gaussian uses its evaluator's phase-center convention, $(p+0.5)0.3/60$, while ours uses dataset phase centers. Mean projection-rendering latency is approximately $23.0\times$ that of $X^2$-Gaussian, and native-grid volume queries are $1.60\times$ slower. Peak allocated memory is lower for our model in both operations.

\par\medskip\noindent
\begin{minipage}[t]{\linewidth}
\captionsetup{type=table}
\centering\footnotesize
\caption{Inference latency and memory on the same RTX 6000 Ada GPU.}
\label{supp:tab:inferencecontrols}
\renewcommand{\arraystretch}{1.08}
\setlength{\tabcolsep}{3pt}
\begin{tabular*}{\linewidth}{@{\extracolsep{\fill}\hspace{\tabcolsep}}lrrrr}
\toprule
& \multicolumn{2}{c}{Latency (ms) $\downarrow$} & \multicolumn{2}{c}{Peak memory (MiB) $\downarrow$}\\
\cmidrule(lr){2-3}\cmidrule(lr){4-5}
Method & Render & Volume & Render & Volume\\
\midrule
$X^2$-Gaussian & \cellcolor[gray]{0.90}\textbf{5.90} & \cellcolor[gray]{0.90}\textbf{209.56} & 2054.30 & 3321.89\\
Ours-Recon & 135.55 & 334.39 & \cellcolor[gray]{0.90}\textbf{1561.62} & \cellcolor[gray]{0.90}\textbf{2927.06}\\
\bottomrule
\end{tabular*}
\par\vspace{3pt}
\begin{minipage}{\linewidth}
\footnotesize Four-case averages of per-case medians. Rendering includes CPU transfer; volume queries remain on the GPU. Memory includes the loaded scene and model.
\end{minipage}
\end{minipage}
\par\medskip

\subsection{Calibration Overhead}
Across all 23 cases, our residual-cache construction averages 41.03\,s,
followed by 2.83\,s for the two-gain fit. The corresponding
$X^2$-Gaussian costs are 2.38 and 3.03\,s. The serialized calibrator
averages 18.76\,MiB for either backbone. 

\subsection{Equal Training Time}
\label{supp:sec:equaltime}
The DIR case 1 control initially permits up to 7,200 training seconds. $X^2$-Gaussian completes its native 30k schedule after 2,563.86\,s, which becomes the training-time allowance for ours. The 0.22-second difference results from stopping after a completed iteration. Both runs start from dataset initialization and seed zero; model initialization, explicit checkpoint I/O, and final evaluation are excluded from the clock. At this budget, ours reaches reported volume PSNR is 1.555\,dB higher, while projection PSNR is approximately 2.00\,dB lower.

\par\medskip\noindent
\begin{minipage}[t]{\linewidth}
\captionsetup{type=table}
\centering\footnotesize
\caption{Equal-time reconstruction on DIR case 1.}
\label{supp:tab:timebudget}
\renewcommand{\arraystretch}{1.08}
\setlength{\tabcolsep}{3pt}
\begin{tabular*}{\linewidth}{@{\extracolsep{\fill}\hspace{\tabcolsep}}lrrr}
\toprule
& & \multicolumn{2}{c}{PSNR (dB) $\uparrow$}\\
\cmidrule(lr){3-4}
Method & Time (s) & Projection & Volume\\
\midrule
$X^2$-Gaussian & 2563.86 & \cellcolor[gray]{0.90}\textbf{44.995} & 44.651\\
Ours (partial) & 2564.08 & 42.990 & \cellcolor[gray]{0.90}\textbf{46.206}\\
\bottomrule
\end{tabular*}
\par\vspace{3pt}
\begin{minipage}{\linewidth}
\footnotesize Both runs start from dataset initialization. Ours stops in the low-frequency stage; projections are uncalibrated.
\end{minipage}
\end{minipage}
\par\medskip

\subsection{Recorded Static-Reconstruction Workloads}
Table~\ref{supp:tab:loggedruntime} lists recorded static reconstruction
durations summed over ten phases per case. Public NAF and SAX-NeRF
schedules entail 11.70 and 5.85 million updates, respectively. The logs
were collected under different devices or loads and describe the
executed workload, rather than a controlled speed comparison.
Historical end-to-end timing for the full dynamic benchmark is unavailable. A public SAX-NeRF runtime
aggregate is not included in the recorded measurements.

\begin{table}[t]
\centering\footnotesize
\caption{Recorded runtime per case (hours; ten phases).
FDK is analytical. Hardware and loads vary across runs.}
\label{supp:tab:loggedruntime}
\renewcommand{\arraystretch}{1.08}
\setlength{\tabcolsep}{2pt}
\begin{tabular*}{\linewidth}{@{\extracolsep{\fill}}llrrr@{}}
\toprule
Benchmark & Method & Cases & Mean (h) & Range (h)\\
\midrule
Public & FDK       & 13 & 0.003 & 0.001--0.006\\
       & R$^2$-GS  & 13 & 2.521 & 1.718--4.120\\
       & NAF       & 13 & 4.074 & 3.147--6.326\\
\midrule
Private & FDK      & 10 & 0.001 & 0.001--0.002\\
        & R$^2$-GS & 10 & 1.530 & 1.445--1.574\\
        & NAF      & 10 & 3.110 & 2.714--3.493\\
        & SAX-NeRF & 10 & 14.42 & 14.00--16.96\\
\bottomrule
\end{tabular*}
\end{table}

\onecolumn
\raggedbottom
\makeatletter
\global\@colht=\textheight
\global\@colroom=\textheight
\global\vsize=\textheight
\makeatother
\noindent\begin{minipage}[t]{\textwidth}
\section{Case-Level Reconstruction Results}
\label{app:cases}
\label{app:allbaselines}
Tables~\ref{supp:tab:allcasesDIR}--\ref{supp:tab:allcasesprivate_volume}
show consistent volumetric advantages over the competing methods.
Ours-Recon exceeds FDK, NAF, and R$^2$-Gaussian in volume PSNR
across all 23 cases, and SAX-NeRF across all ten private cases.
Compared with $X^2$-Gaussian, mean volume-PSNR gains are
1.475, 1.238, and 2.606\,dB on DIR, 4DLung, and SPARE,
respectively, and 0.587\,dB on private data.
Ours-Recon achieves the highest volume PSNR and SSIM among the
tabulated methods in 22 cases; P03 is the exception,
with a volume-PSNR decrease of 0.115\,dB.

Projection and volume rankings differ. On private data,
R$^2$-Gaussian achieves higher uncalibrated projection PSNR
than Ours-Recon but lower volume PSNR.
Calibration improves our projection PSNR across all four subsets.
Ours-Full obtains the highest projection PSNR in all 23 cases
and the highest projection SSIM in 22 cases, except DIR case 5.
On private data, it exceeds identically calibrated
$X^2$-Gaussian by 2.580\,dB while retaining the same
reconstructed volumes as Ours-Recon.

\captionsetup{type=table}
\centering\small
\caption{Per-case reconstruction quality on DIR. All included methods have full subset coverage.}
\label{supp:tab:allcasesDIR}
\renewcommand{\arraystretch}{1.08}
\setlength{\tabcolsep}{4pt}
\begin{tabular*}{\textwidth}{@{\extracolsep{\fill}\hspace{\tabcolsep}}lrrrrrr}
\toprule
Case & FDK & NAF & R$^2$-GS & $X^2$ & Recon & Full\\
\midrule
\multicolumn{7}{l}{\textit{Projection PSNR (dB) $\uparrow$}}\\[2pt]
1 & 24.079 & 39.699 & 43.749 & 45.073 & 44.518 & \cellcolor[gray]{0.90}\textbf{46.747}\\
2 & 22.599 & 42.932 & 43.858 & 44.278 & 44.808 & \cellcolor[gray]{0.90}\textbf{46.950}\\
3 & 22.724 & 40.173 & 43.061 & 41.993 & 41.375 & \cellcolor[gray]{0.90}\textbf{45.206}\\
4 & 23.541 & 42.213 & 43.900 & 44.088 & 43.344 & \cellcolor[gray]{0.90}\textbf{46.608}\\
5 & 23.125 & 42.172 & 44.540 & 44.552 & 44.220 & \cellcolor[gray]{0.90}\textbf{46.284}\\
\addlinespace[2pt]
Mean & 23.214 & 41.438 & 43.822 & 43.997 & 43.653 & 46.359\\
\midrule
\multicolumn{7}{l}{\textit{Projection SSIM $\uparrow$}}\\[2pt]
1 & 0.821352 & 0.902187 & 0.986752 & 0.990458 & 0.990857 & \cellcolor[gray]{0.90}\textbf{0.992204}\\
2 & 0.831181 & 0.972185 & 0.986026 & 0.990445 & 0.990435 & \cellcolor[gray]{0.90}\textbf{0.991609}\\
3 & 0.821254 & 0.921798 & 0.984352 & 0.986222 & 0.984256 & \cellcolor[gray]{0.90}\textbf{0.989027}\\
4 & 0.863473 & 0.945072 & 0.987658 & 0.990400 & 0.990599 & \cellcolor[gray]{0.90}\textbf{0.992070}\\
5 & 0.841759 & 0.971130 & 0.988798 & \cellcolor[gray]{0.90}\textbf{0.993108} & 0.991834 & 0.992999\\
\addlinespace[2pt]
Mean & 0.835804 & 0.942474 & 0.986717 & 0.990127 & 0.989596 & 0.991582\\
\midrule
\multicolumn{7}{l}{\textit{Volume PSNR (dB) $\uparrow$}}\\[2pt]
1 & 31.906 & 40.596 & 42.595 & 44.614 & \cellcolor[gray]{0.90}\textbf{46.835} & \cellcolor[gray]{0.90}\textbf{46.835}\\
2 & 21.612 & 32.304 & 32.217 & 35.328 & \cellcolor[gray]{0.90}\textbf{36.904} & \cellcolor[gray]{0.90}\textbf{36.904}\\
3 & 32.182 & 40.763 & 42.343 & 42.919 & \cellcolor[gray]{0.90}\textbf{44.184} & \cellcolor[gray]{0.90}\textbf{44.184}\\
4 & 25.200 & 34.665 & 35.268 & 37.211 & \cellcolor[gray]{0.90}\textbf{38.817} & \cellcolor[gray]{0.90}\textbf{38.817}\\
5 & 22.358 & 33.306 & 32.924 & 36.367 & \cellcolor[gray]{0.90}\textbf{37.074} & \cellcolor[gray]{0.90}\textbf{37.074}\\
\addlinespace[2pt]
Mean & 26.652 & 36.327 & 37.069 & 39.288 & 40.763 & 40.763\\
\midrule
\multicolumn{7}{l}{\textit{Volume SSIM $\uparrow$}}\\[2pt]
1 & 0.600309 & 0.945081 & 0.963205 & 0.977891 & \cellcolor[gray]{0.90}\textbf{0.985632} & \cellcolor[gray]{0.90}\textbf{0.985632}\\
2 & 0.369060 & 0.844363 & 0.864011 & 0.934783 & \cellcolor[gray]{0.90}\textbf{0.950589} & \cellcolor[gray]{0.90}\textbf{0.950589}\\
3 & 0.658370 & 0.945358 & 0.960434 & 0.969823 & \cellcolor[gray]{0.90}\textbf{0.978810} & \cellcolor[gray]{0.90}\textbf{0.978810}\\
4 & 0.506120 & 0.883643 & 0.904903 & 0.942287 & \cellcolor[gray]{0.90}\textbf{0.958573} & \cellcolor[gray]{0.90}\textbf{0.958573}\\
5 & 0.424955 & 0.870857 & 0.883024 & 0.946705 & \cellcolor[gray]{0.90}\textbf{0.952574} & \cellcolor[gray]{0.90}\textbf{0.952574}\\
\addlinespace[2pt]
Mean & 0.511763 & 0.897860 & 0.915115 & 0.954298 & 0.965236 & 0.965236\\
\bottomrule
\end{tabular*}
\par\vspace{3pt}
\begin{minipage}[t]{\textwidth}
\end{minipage}
\end{minipage}
\par\medskip

\par\medskip\noindent
\begin{minipage}[t]{\textwidth}
\captionsetup{type=table}
\centering\small
\caption{Per-case reconstruction quality on 4DLung. All included methods have full subset coverage.}
\label{supp:tab:allcases4DLung}
\renewcommand{\arraystretch}{1.08}
\setlength{\tabcolsep}{4pt}
\begin{tabular*}{\textwidth}{@{\extracolsep{\fill}\hspace{\tabcolsep}}lrrrrrr}
\toprule
Case & FDK & NAF & R$^2$-GS & $X^2$ & Recon & Full\\
\midrule
\multicolumn{7}{l}{\textit{Projection PSNR (dB) $\uparrow$}}\\[2pt]
1 & 23.045 & 43.207 & 45.895 & 47.035 & 47.716 & \cellcolor[gray]{0.90}\textbf{49.274}\\
2 & 25.358 & 40.238 & 47.662 & 50.189 & 49.631 & \cellcolor[gray]{0.90}\textbf{50.818}\\
3 & 24.659 & 31.188 & 48.434 & 48.479 & 49.365 & \cellcolor[gray]{0.90}\textbf{51.283}\\
4 & 26.110 & 27.438 & 47.589 & 48.443 & 48.867 & \cellcolor[gray]{0.90}\textbf{49.677}\\
5 & 23.886 & 43.644 & 46.301 & 49.306 & 48.647 & \cellcolor[gray]{0.90}\textbf{50.423}\\
\addlinespace[2pt]
Mean & 24.612 & 37.143 & 47.176 & 48.690 & 48.845 & 50.295\\
\midrule
\multicolumn{7}{l}{\textit{Projection SSIM $\uparrow$}}\\[2pt]
1 & 0.689183 & 0.928408 & 0.992266 & 0.995362 & 0.995121 & \cellcolor[gray]{0.90}\textbf{0.995572}\\
2 & 0.763308 & 0.833589 & 0.995142 & 0.996992 & 0.996733 & \cellcolor[gray]{0.90}\textbf{0.997139}\\
3 & 0.765188 & 0.637889 & 0.995384 & 0.996944 & 0.996758 & \cellcolor[gray]{0.90}\textbf{0.997126}\\
4 & 0.803570 & 0.558122 & 0.995057 & 0.996642 & 0.996646 & \cellcolor[gray]{0.90}\textbf{0.996955}\\
5 & 0.710372 & 0.969186 & 0.993830 & 0.996825 & 0.996475 & \cellcolor[gray]{0.90}\textbf{0.996858}\\
\addlinespace[2pt]
Mean & 0.746324 & 0.785439 & 0.994336 & 0.996553 & 0.996346 & 0.996730\\
\midrule
\multicolumn{7}{l}{\textit{Volume PSNR (dB) $\uparrow$}}\\[2pt]
1 & 17.269 & 29.659 & 30.768 & 34.470 & \cellcolor[gray]{0.90}\textbf{35.947} & \cellcolor[gray]{0.90}\textbf{35.947}\\
2 & 22.852 & 34.992 & 36.122 & 40.458 & \cellcolor[gray]{0.90}\textbf{41.755} & \cellcolor[gray]{0.90}\textbf{41.755}\\
3 & 22.673 & 35.280 & 36.215 & 39.954 & \cellcolor[gray]{0.90}\textbf{41.291} & \cellcolor[gray]{0.90}\textbf{41.291}\\
4 & 22.418 & 33.830 & 34.465 & 38.084 & \cellcolor[gray]{0.90}\textbf{39.019} & \cellcolor[gray]{0.90}\textbf{39.019}\\
5 & 21.598 & 33.782 & 34.994 & 40.066 & \cellcolor[gray]{0.90}\textbf{41.208} & \cellcolor[gray]{0.90}\textbf{41.208}\\
\addlinespace[2pt]
Mean & 21.362 & 33.509 & 34.513 & 38.606 & 39.844 & 39.844\\
\midrule
\multicolumn{7}{l}{\textit{Volume SSIM $\uparrow$}}\\[2pt]
1 & 0.123153 & 0.832723 & 0.865993 & 0.929221 & \cellcolor[gray]{0.90}\textbf{0.935790} & \cellcolor[gray]{0.90}\textbf{0.935790}\\
2 & 0.237420 & 0.921552 & 0.940232 & 0.975410 & \cellcolor[gray]{0.90}\textbf{0.979789} & \cellcolor[gray]{0.90}\textbf{0.979789}\\
3 & 0.220351 & 0.911945 & 0.931529 & 0.965661 & \cellcolor[gray]{0.90}\textbf{0.970441} & \cellcolor[gray]{0.90}\textbf{0.970441}\\
4 & 0.232465 & 0.876695 & 0.897236 & 0.942575 & \cellcolor[gray]{0.90}\textbf{0.946722} & \cellcolor[gray]{0.90}\textbf{0.946722}\\
5 & 0.201779 & 0.904330 & 0.923239 & 0.973181 & \cellcolor[gray]{0.90}\textbf{0.977256} & \cellcolor[gray]{0.90}\textbf{0.977256}\\
\addlinespace[2pt]
Mean & 0.203034 & 0.889449 & 0.911646 & 0.957210 & 0.962000 & 0.962000\\
\bottomrule
\end{tabular*}
\end{minipage}

\par\medskip\noindent
\begin{minipage}[t]{\textwidth}
\captionsetup{type=table}
\centering\small
\caption{Per-case reconstruction quality on SPARE. All included methods have full subset coverage.}
\label{supp:tab:allcasesSPARE}
\renewcommand{\arraystretch}{1.08}
\setlength{\tabcolsep}{4pt}
\begin{tabular*}{\textwidth}{@{\extracolsep{\fill}\hspace{\tabcolsep}}lrrrrrr}
\toprule
Case & FDK & NAF & R$^2$-GS & $X^2$ & Recon & Full\\
\midrule
\multicolumn{7}{l}{\textit{Projection PSNR (dB) $\uparrow$}}\\[2pt]
1 & 23.072 & 29.059 & 47.183 & 48.805 & 48.164 & \cellcolor[gray]{0.90}\textbf{52.017}\\
2 & 25.876 & 37.319 & 46.243 & 46.888 & 45.349 & \cellcolor[gray]{0.90}\textbf{50.758}\\
3 & 25.804 & 36.989 & 45.594 & 50.277 & 49.802 & \cellcolor[gray]{0.90}\textbf{52.593}\\
\addlinespace[2pt]
Mean & 24.917 & 34.456 & 46.340 & 48.657 & 47.772 & 51.789\\
\midrule
\multicolumn{7}{l}{\textit{Projection SSIM $\uparrow$}}\\[2pt]
1 & 0.930849 & 0.824258 & 0.994190 & 0.997880 & 0.997340 & \cellcolor[gray]{0.90}\textbf{0.998057}\\
2 & 0.927325 & 0.862880 & 0.994381 & 0.997487 & 0.997142 & \cellcolor[gray]{0.90}\textbf{0.997852}\\
3 & 0.926880 & 0.864381 & 0.993471 & 0.998229 & 0.998030 & \cellcolor[gray]{0.90}\textbf{0.998416}\\
\addlinespace[2pt]
Mean & 0.928352 & 0.850507 & 0.994014 & 0.997865 & 0.997504 & 0.998108\\
\midrule
\multicolumn{7}{l}{\textit{Volume PSNR (dB) $\uparrow$}}\\[2pt]
1 & 14.388 & 26.431 & 26.810 & 31.378 & \cellcolor[gray]{0.90}\textbf{33.627} & \cellcolor[gray]{0.90}\textbf{33.627}\\
2 & 18.352 & 28.411 & 28.899 & 32.578 & \cellcolor[gray]{0.90}\textbf{35.693} & \cellcolor[gray]{0.90}\textbf{35.693}\\
3 & 17.597 & 26.948 & 27.340 & 32.895 & \cellcolor[gray]{0.90}\textbf{35.349} & \cellcolor[gray]{0.90}\textbf{35.349}\\
\addlinespace[2pt]
Mean & 16.779 & 27.263 & 27.683 & 32.284 & 34.890 & 34.890\\
\midrule
\multicolumn{7}{l}{\textit{Volume SSIM $\uparrow$}}\\[2pt]
1 & 0.425490 & 0.785814 & 0.827724 & 0.919965 & \cellcolor[gray]{0.90}\textbf{0.933606} & \cellcolor[gray]{0.90}\textbf{0.933606}\\
2 & 0.362670 & 0.808841 & 0.857963 & 0.907628 & \cellcolor[gray]{0.90}\textbf{0.925370} & \cellcolor[gray]{0.90}\textbf{0.925370}\\
3 & 0.380939 & 0.776869 & 0.834656 & 0.938532 & \cellcolor[gray]{0.90}\textbf{0.951513} & \cellcolor[gray]{0.90}\textbf{0.951513}\\
\addlinespace[2pt]
Mean & 0.389700 & 0.790508 & 0.840114 & 0.922041 & 0.936830 & 0.936830\\
\bottomrule
\end{tabular*}
\end{minipage}
\par\medskip

\par\medskip\noindent
\begin{minipage}[t]{\textwidth}
\captionsetup{type=table}
\centering\footnotesize
\caption{Private case-level projection metrics. All included methods have ten-case coverage.}
\label{supp:tab:allcasesprivate_projection}
\renewcommand{\arraystretch}{1.08}
\setlength{\tabcolsep}{3pt}
\begin{tabular*}{\textwidth}{@{\extracolsep{\fill}\hspace{\tabcolsep}}lrrrrrrrr}
\toprule
Case & FDK & NAF & SAX-NeRF & R$^2$-GS & $X^2$ & $X^2$+Cal. & Recon & Full\\
\midrule
\multicolumn{9}{l}{\textit{Projection PSNR (dB) $\uparrow$}}\\[2pt]
P01 & 23.389 & 45.890 & 46.001 & 50.377 & 51.105 & 52.251 & 51.027 & \cellcolor[gray]{0.90}\textbf{52.666}\\
P02 & 23.985 & 24.331 & 47.288 & 51.925 & 49.261 & 50.379 & 48.325 & \cellcolor[gray]{0.90}\textbf{53.132}\\
P03 & 22.796 & 35.127 & 46.228 & 50.221 & 46.038 & 46.986 & 45.604 & \cellcolor[gray]{0.90}\textbf{50.803}\\
P04 & 24.189 & 28.178 & 45.894 & 50.687 & 50.526 & 51.484 & 50.772 & \cellcolor[gray]{0.90}\textbf{53.530}\\
P05 & 23.193 & 32.435 & 45.901 & 50.560 & 46.853 & 47.350 & 46.547 & \cellcolor[gray]{0.90}\textbf{51.327}\\
P06 & 20.901 & 42.778 & 42.762 & 47.496 & 48.762 & 49.165 & 48.895 & \cellcolor[gray]{0.90}\textbf{50.899}\\
P07 & 23.274 & 27.492 & 46.885 & 51.110 & 48.998 & 49.787 & 49.291 & \cellcolor[gray]{0.90}\textbf{52.746}\\
P08 & 24.893 & 27.228 & 48.252 & 52.468 & 51.131 & 51.627 & 52.135 & \cellcolor[gray]{0.90}\textbf{54.769}\\
P09 & 24.805 & 25.595 & 47.041 & 51.553 & 50.989 & 51.769 & 50.815 & \cellcolor[gray]{0.90}\textbf{53.733}\\
P10 & 23.262 & 41.041 & 47.281 & 51.259 & 48.973 & 49.626 & 49.097 & \cellcolor[gray]{0.90}\textbf{52.619}\\
\addlinespace[2pt]
Mean & 23.469 & 33.010 & 46.353 & 50.766 & 49.264 & 50.042 & 49.251 & 52.622\\
\midrule
\multicolumn{9}{l}{\textit{Projection SSIM $\uparrow$}}\\[2pt]
P01 & 0.634213 & 0.974851 & 0.988902 & 0.995934 & 0.996079 & 0.996846 & 0.995157 & \cellcolor[gray]{0.90}\textbf{0.997060}\\
P02 & 0.758048 & 0.597943 & 0.990749 & 0.997077 & 0.996597 & 0.997224 & 0.996221 & \cellcolor[gray]{0.90}\textbf{0.997817}\\
P03 & 0.728393 & 0.820487 & 0.989069 & 0.995855 & 0.994652 & 0.995499 & 0.994162 & \cellcolor[gray]{0.90}\textbf{0.996773}\\
P04 & 0.724487 & 0.663906 & 0.988619 & 0.996617 & 0.996505 & 0.997317 & 0.996363 & \cellcolor[gray]{0.90}\textbf{0.997715}\\
P05 & 0.705099 & 0.760030 & 0.989225 & 0.995953 & 0.995281 & 0.995813 & 0.994883 & \cellcolor[gray]{0.90}\textbf{0.996868}\\
P06 & 0.636299 & 0.964645 & 0.980900 & 0.993948 & 0.994498 & 0.995246 & 0.993718 & \cellcolor[gray]{0.90}\textbf{0.996029}\\
P07 & 0.737885 & 0.679650 & 0.989978 & 0.996417 & 0.996110 & 0.996708 & 0.996169 & \cellcolor[gray]{0.90}\textbf{0.997436}\\
P08 & 0.757457 & 0.648948 & 0.992144 & 0.997607 & 0.997639 & 0.997892 & 0.997678 & \cellcolor[gray]{0.90}\textbf{0.998420}\\
P09 & 0.764890 & 0.602712 & 0.991404 & 0.997125 & 0.997076 & 0.997544 & 0.996914 & \cellcolor[gray]{0.90}\textbf{0.997915}\\
P10 & 0.661450 & 0.885321 & 0.989760 & 0.996438 & 0.996022 & 0.996525 & 0.995546 & \cellcolor[gray]{0.90}\textbf{0.997225}\\
\addlinespace[2pt]
Mean & 0.710822 & 0.759849 & 0.989075 & 0.996297 & 0.996046 & 0.996662 & 0.995681 & 0.997326\\
\bottomrule
\end{tabular*}
\end{minipage}

\par\medskip\noindent
\begin{minipage}[t]{\textwidth}
\captionsetup{type=table}
\centering\footnotesize
\caption{Private case-level volume metrics. All included methods have ten-case coverage.}
\label{supp:tab:allcasesprivate_volume}
\renewcommand{\arraystretch}{1.08}
\setlength{\tabcolsep}{3pt}
\begin{tabular*}{\textwidth}{@{\extracolsep{\fill}\hspace{\tabcolsep}}lrrrrrrrr}
\toprule
Case & FDK & NAF & SAX-NeRF & R$^2$-GS & $X^2$ & $X^2$+Cal. & Recon & Full\\
\midrule
\multicolumn{9}{l}{\textit{Volume PSNR (dB) $\uparrow$}}\\[2pt]
P01 & 31.892 & 45.203 & 46.050 & 47.830 & 52.479 & 52.479 & \cellcolor[gray]{0.90}\textbf{53.260} & \cellcolor[gray]{0.90}\textbf{53.260}\\
P02 & 31.171 & 43.187 & 43.922 & 45.467 & 48.600 & 48.600 & \cellcolor[gray]{0.90}\textbf{48.869} & \cellcolor[gray]{0.90}\textbf{48.869}\\
P03 & 31.629 & 43.678 & 44.565 & 46.221 & \cellcolor[gray]{0.90}\textbf{48.391} & \cellcolor[gray]{0.90}\textbf{48.391} & 48.276 & 48.276\\
P04 & 31.583 & 43.369 & 44.019 & 46.177 & 50.811 & 50.811 & \cellcolor[gray]{0.90}\textbf{51.445} & \cellcolor[gray]{0.90}\textbf{51.445}\\
P05 & 30.821 & 43.267 & 43.989 & 45.349 & 48.383 & 48.383 & \cellcolor[gray]{0.90}\textbf{48.737} & \cellcolor[gray]{0.90}\textbf{48.737}\\
P06 & 31.672 & 44.676 & 45.449 & 47.143 & 50.048 & 50.048 & \cellcolor[gray]{0.90}\textbf{51.334} & \cellcolor[gray]{0.90}\textbf{51.334}\\
P07 & 30.807 & 42.977 & 43.661 & 45.106 & 48.414 & 48.414 & \cellcolor[gray]{0.90}\textbf{49.001} & \cellcolor[gray]{0.90}\textbf{49.001}\\
P08 & 32.313 & 45.388 & 46.065 & 47.464 & 50.456 & 50.456 & \cellcolor[gray]{0.90}\textbf{51.282} & \cellcolor[gray]{0.90}\textbf{51.282}\\
P09 & 32.284 & 44.304 & 45.011 & 46.617 & 49.879 & 49.879 & \cellcolor[gray]{0.90}\textbf{50.502} & \cellcolor[gray]{0.90}\textbf{50.502}\\
P10 & 31.708 & 45.476 & 46.206 & 47.585 & 49.977 & 49.977 & \cellcolor[gray]{0.90}\textbf{50.604} & \cellcolor[gray]{0.90}\textbf{50.604}\\
\addlinespace[2pt]
Mean & 31.588 & 44.153 & 44.894 & 46.496 & 49.744 & 49.744 & 50.331 & 50.331\\
\midrule
\multicolumn{9}{l}{\textit{Volume SSIM $\uparrow$}}\\[2pt]
P01 & 0.476689 & 0.976247 & 0.980210 & 0.987206 & 0.995872 & 0.995872 & \cellcolor[gray]{0.90}\textbf{0.996350} & \cellcolor[gray]{0.90}\textbf{0.996350}\\
P02 & 0.463440 & 0.966875 & 0.970504 & 0.978807 & 0.991593 & 0.991593 & \cellcolor[gray]{0.90}\textbf{0.991626} & \cellcolor[gray]{0.90}\textbf{0.991626}\\
P03 & 0.491938 & 0.969002 & 0.973301 & 0.981445 & \cellcolor[gray]{0.90}\textbf{0.990694} & \cellcolor[gray]{0.90}\textbf{0.990694} & 0.990254 & 0.990254\\
P04 & 0.470989 & 0.968045 & 0.971677 & 0.981185 & 0.994094 & 0.994094 & \cellcolor[gray]{0.90}\textbf{0.994837} & \cellcolor[gray]{0.90}\textbf{0.994837}\\
P05 & 0.444268 & 0.966976 & 0.970788 & 0.977670 & 0.990451 & 0.990451 & \cellcolor[gray]{0.90}\textbf{0.991107} & \cellcolor[gray]{0.90}\textbf{0.991107}\\
P06 & 0.482758 & 0.958283 & 0.964267 & 0.976434 & 0.986251 & 0.986251 & \cellcolor[gray]{0.90}\textbf{0.988335} & \cellcolor[gray]{0.90}\textbf{0.988335}\\
P07 & 0.453499 & 0.964828 & 0.968583 & 0.977743 & 0.991385 & 0.991385 & \cellcolor[gray]{0.90}\textbf{0.991723} & \cellcolor[gray]{0.90}\textbf{0.991723}\\
P08 & 0.499700 & 0.976396 & 0.979117 & 0.984786 & 0.993905 & 0.993905 & \cellcolor[gray]{0.90}\textbf{0.994002} & \cellcolor[gray]{0.90}\textbf{0.994002}\\
P09 & 0.500470 & 0.970698 & 0.973768 & 0.982386 & 0.993769 & 0.993769 & \cellcolor[gray]{0.90}\textbf{0.993857} & \cellcolor[gray]{0.90}\textbf{0.993857}\\
P10 & 0.460631 & 0.979249 & 0.981610 & 0.986785 & 0.993838 & 0.993838 & \cellcolor[gray]{0.90}\textbf{0.993961} & \cellcolor[gray]{0.90}\textbf{0.993961}\\
\addlinespace[2pt]
Mean & 0.474438 & 0.969660 & 0.973383 & 0.981445 & 0.992185 & 0.992185 & 0.992605 & 0.992605\\
\bottomrule
\end{tabular*}
\end{minipage}

\section{Additional Qualitative Comparisons}
\label{app:qualitative}

\subsection{Additional Volume Examples}
Figures~\ref{supp:fig:qual_dir1} complement
the main-paper comparison with full-resolution panels. DIR case 1
includes an available SAX-NeRF phase. SPARE case 2 has the largest
public case-mean volume-PSNR gain over $X^2$-Gaussian, whereas private
P09 is the lower-middle private case under the same ranking. P06 phase
9 is the strongest private phase-level example. These selection rules
describe the illustrations; the complete available case tables are in
Sec.~\ref{app:cases}.

\par\medskip\noindent
\begin{minipage}[t]{\textwidth}
\captionsetup{type=figure}
\centering
\includegraphics[width=0.8 \textwidth]{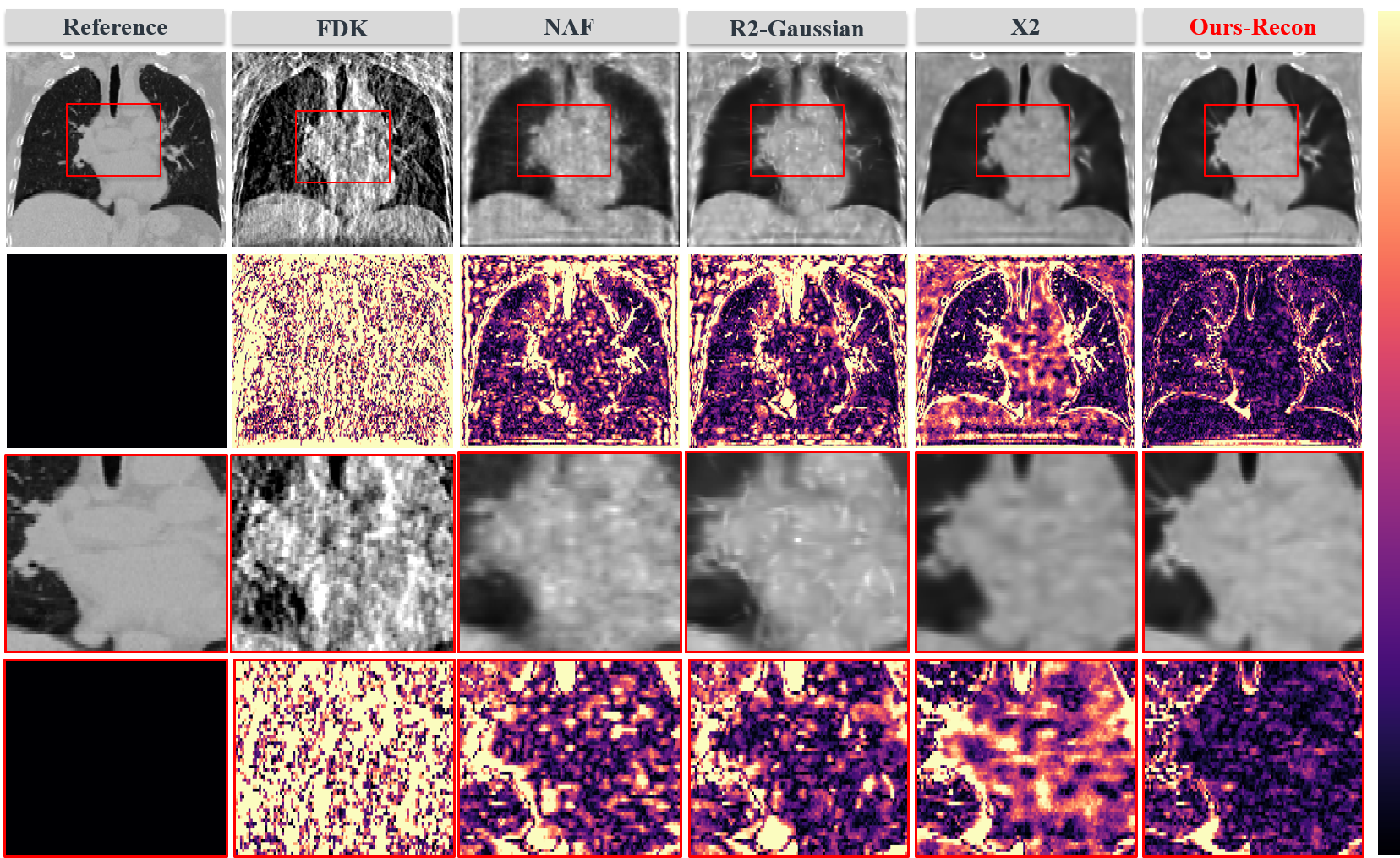}
\caption{DIR case 1, phase 0. Rows show the central slice, absolute error, fixed crop, and crop error. SAX-NeRF is available for this phase. The Ours-Recon volume also represents Ours-Full.}
\label{supp:fig:qual_dir1}
\end{minipage}
\par\medskip



\subsection{Quality at Observed Respiratory Phases}
\label{supp:sec:consistency}
Figure~\ref{supp:fig:phaseprofiles} reports full-volume PSNR at the ten observed respiratory phase centers for six illustrated cases. These curves reveal within-cycle variation that is hidden by case-averaged results. FDK consistently yields lower PSNR, while NAF, SAX-NeRF, and R$^2$-Gaussian improve reconstruction quality. The two dynamic methods generally achieve higher volumetric fidelity. Compared with $X^2$-Gaussian, Ours-Recon shows clear improvements across the observed phases of DIR case 1, 4DLung case 1, and SPARE case 2. In SPARE case 2, several baselines exhibit a pronounced quality drop near phase 4, whereas Ours-Recon maintains higher PSNR with a smaller decline. The two private cases show smaller but consistent gains across phases. However, curve crossings in DIR case 3 indicate that Ours-Recon does not outperform every competing method at every phase. These results characterize the final reconstructions at observed phase centers. The effect of phase-isolated updates is evaluated separately by the controlled recovery experiments.

\par\medskip\noindent
\begin{minipage}[t]{\textwidth}
\captionsetup{type=figure}
\centering
\includegraphics[width=1.0 \textwidth]{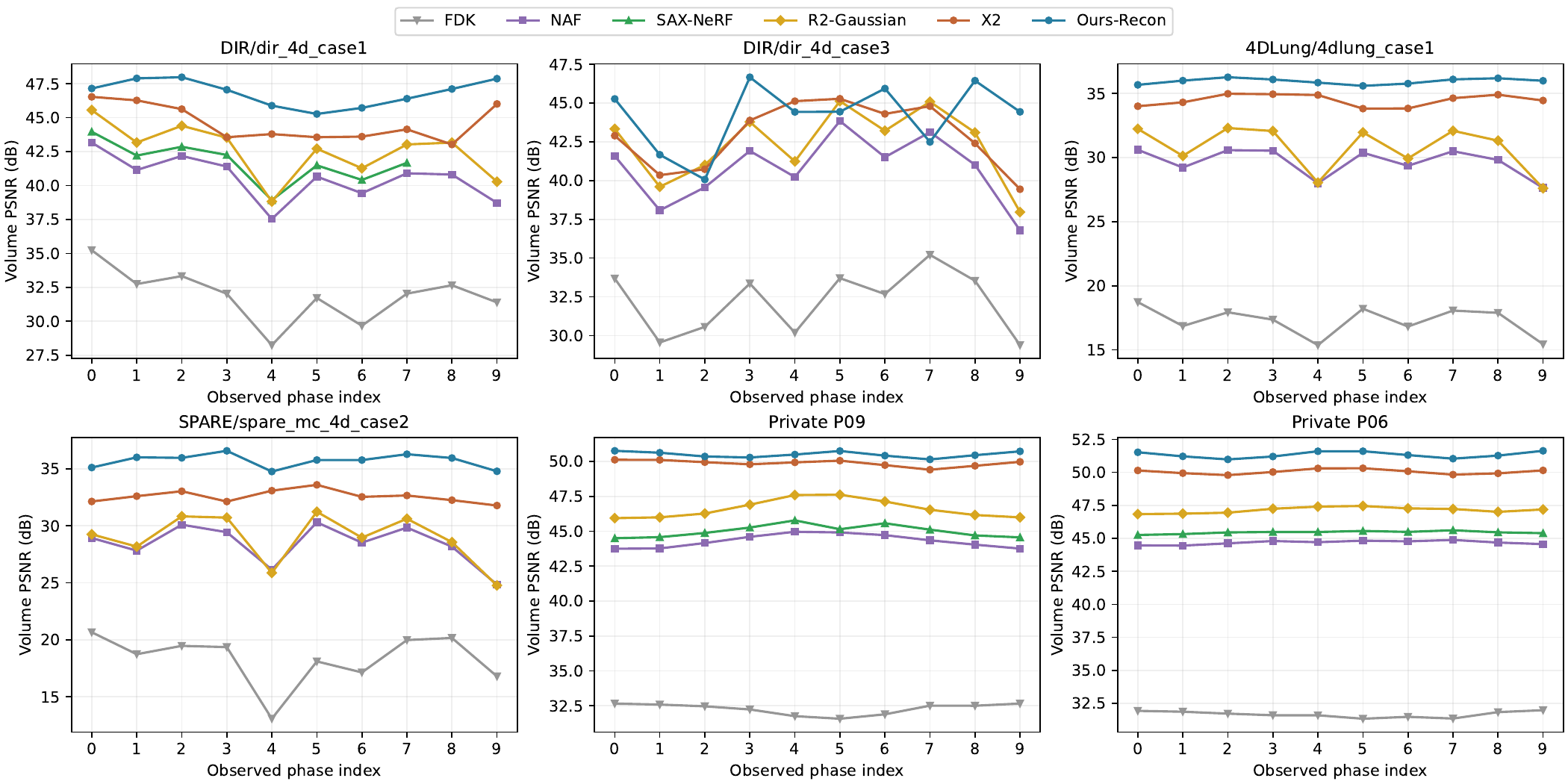}
\caption{Full-volume PSNR at the ten reference phase centers for six illustrated cases. The plot shows available phase-level SAX-NeRF records; missing phase traces are left unplotted.}
\label{supp:fig:phaseprofiles}
\end{minipage}
\par\medskip

\subsection{Held-Out Projection Examples}
The error maps reveal differences that are less apparent in the projection images. FDK exhibits extensive residuals, while NAF retains pronounced boundary errors in the upper example. Gaussian-based methods reduce these errors, although their relative performance varies across views. In the lower example, Ours-Recon shows a bright residual band near the lung base, which is markedly
reduced by Ours-Full. This illustrates the benefit of projection calibration for localized discrepancies while keeping the reconstructed volume fixed.

\par\medskip\noindent
\begin{minipage}[t]{\textwidth}
\captionsetup{type=figure}
\centering
\includegraphics[width=0.8 \textwidth]{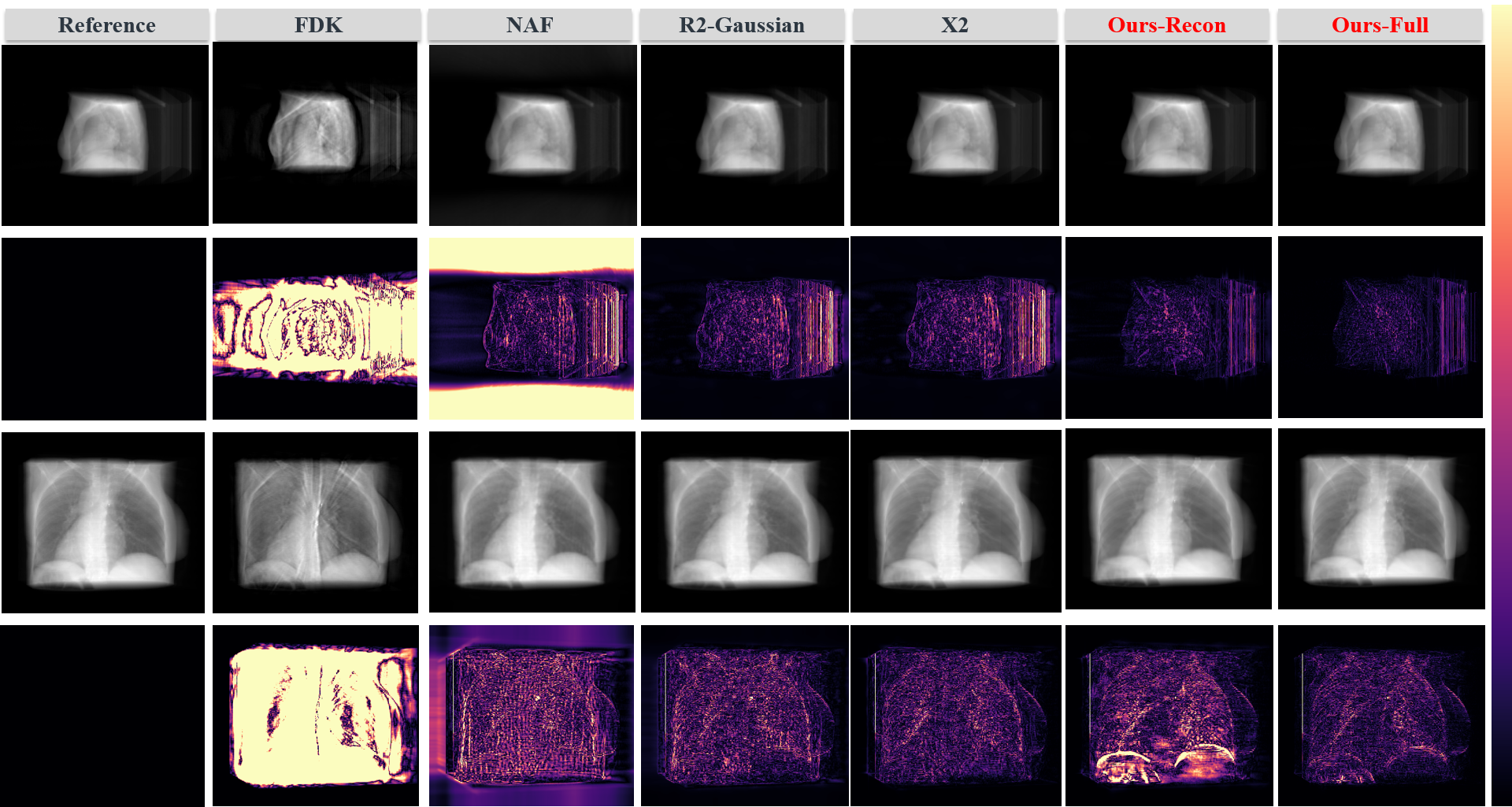}
\caption{First held-out phase-0 view of DIR case 1. Rows show normalized projections and shared-scale absolute errors. SAX-NeRF is available for this phase. Ours-Recon and Ours-Full use the same volume.}
\label{supp:fig:projectionpublic}
\end{minipage}
\par\medskip


{
    \small
    \bibliographystyle{ieeenat_fullname}
    \bibliography{main_2}
}

\end{document}